\documentclass[%
reprint,
amsmath,amssymb,
aps,
]{revtex4-1}
\usepackage{graphicx}
\usepackage{amsthm,amsmath}
\usepackage[top=25.4mm, bottom=20.9mm, left=19.1mm, right=19.1mm]{geometry}
\usepackage{mathtools}
\DeclarePairedDelimiter\ceil{\lceil}{\rceil}
\DeclarePairedDelimiter\floor{\lfloor}{\rfloor}
\theoremstyle{definition} 

\newcommand{\DD}{\, \displaystyle}
\newcommand{\deff}{\, \stackrel{\text{def}}{=}}

\newcommand{\rond}[2]{\, \frac{\partial #1}{\partial #2}}
\newcommand{\rondd}[2]{\, \frac{\partial^2#1}{\partial #2^2}}
\newcommand{\fracc}[2]{\, \displaystyle \frac{ #1}{ #2}}

\newcommand{\morabba}[1]{\,\begin{flushright}
\Rectsteel \\
\end{flushright}}

\newcommand{\CC}[2]{\, \binom{#1} {#2} 
}

\newcommand{\al}[1]{\,\begin{align}
#1 
\end{align}
}
\newcommand{\all}[2]{\,\begin{align}
#1 
\label{#2}
\end{align}
}

\makeatletter
\newcommand{\vast}{\bBigg@{4}}
\newcommand{\Vast}{\bBigg@{5}}

\makeatother

\begin{document}
\preprint{APS/123-QED}
\title{Temporal Dynamics of Connectivity and Epidemic Properties of Growing Networks}

\author{ Babak Fotouhi$^{1,2,\dagger}$ and Mehrdad Khani Shirkoohi$^{3,4}$ \\
\emph{
$^1$Clinical and Health Informatics Group, 
McGill University, Montr\'eal, Qu\'ebec, Canada\\
$^{2}$Department of Sociology, McGill University, Montr\'eal, Qu\'ebec, Canada}\\
$^{\dagger}$Email:\texttt{ babak.fotouhi@mail.mcgill.ca }\\
\emph{$^3$ Department of Electrical Engineering, Sharif University of Technology, Tehran, Iran}
\\
\emph{$^4$ Department of Computer Science, Sharif University of Technology, Tehran, Iran}
}


\begin{abstract}
Traditional mathematical models of epidemic disease had  for decades conventionally considered static structure for contacts. Recently, an upsurge of theoretical inquiry has strived towards rendering the models more realistic by incorporating the temporal aspects of networks of contacts, societal and online, that are of interest in the study of epidemics (and other similar diffusion processes). However, temporal dynamics have predominantly focused on link fluctuations and nodal activities, and less attention has been paid to the growth of the underlying network. Many real networks grow: online networks are evidently in constant growth, and societal networks can grow due to migration flux and reproduction. The effect of network growth on the epidemic properties of networks  is hitherto unknown---mainly due to the predominant focus of the network growth literature on the so-called steady-state. 
This paper takes a step towards alleviating this gap. We   analytically study the degree dynamics of a given arbitrary network that is subject to growth. We   use the theoretical findings to predict the epidemic properties of the network as a function of time. We observe that the introduction of new individuals into the network can enhance or diminish its resilience against endemic outbreaks, and investigate how this regime shift depends upon the connectivity of newcomers and on how they establish  connections to existing nodes. 
Throughout, theoretical findings are corroborated with Monte Carlo simulations over synthetic and real networks. The results shed light on the effects of network growth on the future epidemic properties of networks, and offers   insights for devising a-priori immunization strategies. 
%
%
%
\end{abstract}

\maketitle


\section{Introduction} \label{sec:intro}

\newcommand{\ave}[1]{\, \langle{#1}\rangle}
Mathematical models of infectious disease, now over  two centuries old~\cite{hethcote2000mathematics,RevModPhys.87.925}, seek to quantify the spread of a disease and aim to predict its prevalence. In their inchoate forms, they assumed an all-to-all (\emph{well-mixed}) contact pattern for the individuals in the population. Attention to the structure of connections is a comparatively recent development, with myriads of studies from different disciplines  investigating  the effects of network structure on disease dynamics~\cite{RevModPhys.87.925}. Most of these analyses  have predominantly focused on long-time predictions. Recent advances in network epidemiology are seeking to alleviate this shortcoming along two main theoretical directions. The first is extending the analysis beyond the confines of the steady-state, and focusing on the dynamics at intermediary  or short time regimes. 
Recent examples of studies that undertake this approach include studying the time evolution of predictability of the eventual outbreak size of an epidemic as a function of the time passed since its inception~\cite{holme2015time}, 
the time-dependent survival probability of the outbreak after an observation at a given time~\cite{holme2015shadows}, the feasibility of quarantine strategies   as a function of the time passed since the inception of  an  epidemic~\cite{pereira2015control}, adaptive immunization strategies that are continuously optimized based on the local hitherto-available information on the history of epidemic outbreak cycles~\cite{yan2014dynamical}, and estimating the past states and the source of the spread based on observations at later times~\cite{altarelli2014bayesian,antulov2015identification}. The second major theme of research (which also subsumes the present paper) incorporates the time-dependence of the medium through which the pathogen spreads, namely, the web of connections. Examples of approaches within this strand of research include incorporating into the models temporal fluctuations of contacts and burstiness of nodal activities and inetractions~\cite{volz2007susceptible,masuda2013predicting,zhang2014susceptible,jo2014analytically,valdano2015analytical} (also see~\cite{holme2015modern}), studying how the epidemic dynamics and thresholds are affected by, for instance, distribution of enter-event times~\cite{vazquez2007impact,perra2012activity,starnini2014temporal,holme2015basic}, strength of ties~\cite{karsai2014time}, and heterogeneous link lifetimes~\cite{sunny2015dynamics}, as well as adaptive models in which the structure of the contact network adapts to the epidemic process~\cite{funk2010modelling,lagorio2011quarantine,fenichel2011adaptive}---that is, the social ties evolve in response to the infection status of nodes. 
The said studies do envisage dynamism for the network structure, but   with fixed size. That is, links do fluctuate or rewire, but number of nodes do not vary. However, many real networks grow in size. Almost all social networks do. There is a paucity of research on the effects of network growth an its concomitant structural changes in  the epidemic behavior of the system. The present paper takes a step towards alleviating this gap. 

The peaks of epidemic cycles for different diseases can be several years apart e.g., 8-10 years for syphilis~\cite{grassly2005host}, 8 years for hepatitis A~\cite{xu2008decline}, $<5$ years for pertussis~\cite{gomes1999diphtheria,sharp2012chikungunya} and measles ~\cite{stone2007seasonal,mantilla2010decreasing},~$\simeq5$ years for mumps~\cite{rohani2010never} and meningitis~\cite{levine2014time}.
In the meantime, populations can grow due to reproduction and migration. This can alter the network structure---the smaller the community, the more substantial the potential change can be. 
Due to the obvious imperativeness of intervention, it would be highly beneficial to utilize the data from the past in order to estimate the epidemic properties of the present. Moreover, sexual contacts networks grow in size and there is evidence that  their  growth mechanism has preferential-attachment characteristics~\cite{liljeros2001web}. Given data on their current and past size and structure, it would be highly valuable if we could theoretically provide predictions for the future epidemic properties of the network.

Motivations for extending epidemic models to growing networks also exist beyond the realm of epidemiology. The versatility of basic compartmental models of epidemic spread have led researchers to employ them in modeling diffusion processes other than the spread of pathogen. For example, the mathematical models of epidemic diseases shares similarities with those employed for the studying of the diffusion of malware, information and emotions on online networks~\cite{hill2010emotions,hill2010infectious,cha2008characterizing,zhao2011rumor}. Note that all these networks are growing networks. 
Similarly, the WWW is also growing, and the study of the spread of computer and mobile viruses and malware on the web is also mathematically akin to epidemic diseases, hence they can be formulated under the same overarching framework~\cite{cohen2003efficient,pastor_book_ch6,pastor2001epidemic,draief2006epidemic,valler2011epidemic,yang2014new}. 
The diversity  of  these potential applications motivates   the extension  the study of epidemic  models  to growing networks. In this paper, we take a first step towards this aim. We need to first analyze the structure of growing networks theoretically, and calculate how the structural properties of a given (arbitrary) network evolves over time. We employ a  basic  model that exists in the network growth literature, as discussed below.


The network growth literature was initiated (or, considering~\cite{cad1976generai}, revived) by the seminal Barab{\'a}si-Albert (henceforth BA) model posited initially in~\cite{barabasi1999emergence}. The main motivation of network growth studies has been to link micro processes that underlie the growth mechanism to macro properties of the network (such as a power-law degree distribution in the case of the BA model). 
Examples of other growth models include models with edge growth~\cite{albert2000topology}, aging effects~\cite{dorogovtsev2000evolution}, node deletion~\cite{moore2006exact}, 
accelerated growth~\cite{dorogovtsev2000effect},
copying~\cite{krapivsky2000connectivity}, as well as fitness-based models~\cite{bianconi2001competition,smolyarenko2013network}.


In all the examples mentioned above, the analysis is confined to the steady-state limit, that is, the limit as $t \rightarrow \infty$. The initial network is conventionally a small one whose effects vanish in the steady state. Thus, the available theoretical results are not suitable for the purposes of the present paper. The minimum that is required for our purposes   is 
a time-dependent  solution (at arbitrary times, not necessarily large times) for the degree distribution of an arbitrary network (with arbitrary topology and size, not necessarily small). This would  constitute  the minimum required information to conduct a basic  heterogeneous mean-field analysis to conventional epidemic models~\cite{RevModPhys.87.925}. For more rigorous analysis, one would require further temporal solutions (e.g., time-dependent degree correlations). In this paper, we only consider the basic setup, and we focus on the degree distribution.

For the growth mechanism, we consider the case of preferential growth with initial attractiveness, in which the growth kernel is linear with a constant shift~\cite{dorogovtsev2000structure}. This model is versatile for subsuming growth processes with diverse underlying mechanisms. The basic version of the model can be interpreted as simple preferential attachment where each node is endowed with an initial chance of receiving links~\cite{krapivsky2001organization}. Furthermore, the local version of the preferential attachment, which does not require knowledge of the global network topology for incoming nodes---for which each new node first chooses an existing node randomly and then redirects its links to the neighbors of the target node---can be reformulated as a shifted-linear model. Moreover, the directed version of the preferential attachment growth---in which the likelihood of each node to receive a link from newcomers linearly depends on both  its  in-degree and out-degree---can be expressed as a shifted-linear growth problem~\cite{krapivsky2001organization}. 

The shifted-linear preferential growth model has the advantage that  it can interpolate between preferential attachment and uniform growth. As shall be discussed in the text, in this model, each new node that is added to the network forms new connections to existing nodes, and the probability that an incoming node with degree $k$ receives a link is proportional to $k+\theta$, where $\theta$ is the  initial attractiveness. For large values of $\theta$, the influence of the  preferential part of the connection kernel is reduced, and the growth mechanism becomes agnostic to the degrees of the existing nodes. On the other extreme, if $\theta=0$, we recover the conventional BA model. 

The rest of this paper is organized as follows. First, we introduce the growth mechanism and quantify the evolution of the quantities of interest via a rate equation approach. We solve the resulting difference-differential equation and find a closed-form solution for the degree distribution as a function of time for arbitrary initial conditions---which, as a byproduct of this paper, can be fruitful contribution for studying diverse processes on growing networks. We then employ the solution to analyze  the   SIR and SIS models of epidemic spread on top of various topologies.  The results indicate that the future epidemic properties of a network can change dramatically---even in intermediary time regimes---depending on the value of the initial attractiveness and on the number of initial connections that each incoming node establishes.  Note that throughout the paper, the calculated epidemic properties are instantaneous, that is, they pertain to \emph{potential} epidemic outbreaks. We are not considering the  problem where the spread of disease is concurrent with network growth, where the two dynamic processes would influence one another.   In the problem that we focus on, the  assumption is  that the growth process is much slower than the time scales of the epidemic disease (e.g.,  seasonal flu epidemics  as compared to population growth). This is effectively an adiabatic approximation, and implies that the growth process is slow enough  so that the network size can be assumed constant during the disease lifetime.  Studying the evolution of such   instantaneous  properties can help us characterize the variation of the susceptibility of the network against potential disease outbreaks at different times.  Throughout, we corroborate the theoretical results with Monte Carlo simulations. Theoretical predictions are in good agreement with simulation results.


\section{Growth Model}

The growth process starts from a given initial network with $N(0)$ nodes and $L(0)$ links, with known degree distribution $p_k(0)$. The network grows via the successive addition of new nodes. At each time step a new node is born, and it forms $\beta$ links to existing nodes in the following way: the probability that an existing node $x$ receives a link from the new node at timestep $n$ is proportional to $k_x(n)+\theta$, where $k_x(n)$ is the degree of node $x$ at timestep $n$, and $\theta$ is the initial attractiveness, a positive constant of the model. To obtain normalized probabilities, we need to divide $k_x(n)+\theta$ for each $x$ by the sum of this quantity over every node. The sum over $k_x$ yields twice the number of links at timestep $n$, which is $2L(0)+2\beta n$. The sum over $\theta$ yields $\theta N(n)$, where $N(n)$ is the number of nodes at timestep $n$, which equals $N(0)+n$. Thus, the probability that node $x$ receives a link emanated from the newly-born node equals 
\all{
\pi_x(n)=\frac{k_x+\theta}{2L(0)+2\beta n+\theta N(0) + \theta n}
.}{pi_x_0}
Hereinafter, we will denote $2L(0)+\theta N(0)$ by $\zeta$, and $2\beta+\theta$ by $\nu$. So~\eqref{pi_x_0} transforms into
\all{
\pi_x(n)=\frac{k_x+\theta}{\zeta + \nu n}
.}{pi_x}


\section{Evolution of the Degrees}
At each timestep, we can quantify the expected   change in $N_k(n)$, which is the number of nodes in the network that have degree $k$ at timestep $n$. The value of $N_k(n)$ can be altered if at time $n$, an existing node with degree $k$ receives a link from the newly-born node (which would increment the degree of the receiving node to $k+1$, decrementing $N_k$), or if an existing node with degree $k-1$ receives a link (which would increment the degree of the receiving node to $k$, incrementing $N_k$). For each incoming node, $N_\beta (n)$ increments. The following rate equation quantifies the evolution of $N_k(n)$: 

\all{
&N_k(n+1)-N_k(n)= \nonumber \\ &
\beta \left[\fracc{(k-1+\theta)N_{k-1}(n)-(k+\theta)N_k(n)}{\zeta + \nu n} \right]
+ \delta_{k,\beta}.
}{Nk_dot_0}

%
%

This is a two-dimensional difference equation in $n$ and $k$. Let us consider the time-continuous analog of $N_k(n)$, and denote it by $M_k(t)$. The values of   $M_k(t)$ at discrete time points $t=n$ is intended to be a good approximation for $N_k(n)$. At the outset, we   have ${M_k(0)=N_k(0), \forall k}$. Also, in the limit at $t \rightarrow \infty$, we require that the ratio $\frac{M_k(n)}{N_k(n)}$ approach unity. Finally, we require that for every $n>0$, the error be reasonably small. We define the error as $E_k(n) \stackrel{\text{def}}{=} N_k(n)-M_k(n)$. The first step we undertake is solving the following difference-differential equation, which is the time-continuous analog of~\eqref{Nk_dot_0}:

\all{
\rond{M_k(t)}{t}&= \beta \left[\fracc{(k-1+\theta)M_{k-1}(t)-(k+\theta)M_k(t)}{\zeta + \nu t} \right]
+ \delta_{k,\beta}.
}{Nk_dot}

After solving this equation, we will investigate the behavior of error $E_k(n)$ both analytically and via simulations, and show that the error is remarkably small, which evinces the  high  accuracy of the approximation employed.
%

To solve~\eqref{Nk_dot}, we define the generating function: 
\all{
\psi(z,t) \stackrel{\text{def}}{=} \DD \sum_{k=1}^{\infty} M_k(t) z^{-k}
.}{psi_def}

This is the conventional Z transform. We multiply both sides of~\eqref{Nk_dot} by $z^{-k}$ and sum over $k$. The left hand side yields $\rond{\psi}{t}$. For the terms on the right hand side, we use two standard properties of the Z-transform: if the generation function of some sequence $a_k$ is given by $A(z)$, then (1) the generating function for sequence $ka_k$ is given by $-z \frac{d A(z)}{dz}$, and (2) the generating function for the sequence $a_{k-1}$ is given by $z^{-1}a_k$. Using these two properties, Equation~\eqref{Nk_dot} yields
\all{
&\DD \rond{\psi(z,t)}{t} \nonumber \\ &
= \fracc{\beta }{\zeta + \nu t} \Bigg[ (z-1) \rond{\psi(z,t)}{z} + \theta (z^{-1}-1) \psi(z,t) \Bigg] + z^{-\beta}
.}{psi_dot_1}
This can be rearranged and recast as
\all{
& \DD \rond{\psi(z,t)}{t} - \fracc{\beta }{\zeta + \nu t} (z-1) \rond{\psi(z,t)}{z} 
\nonumber \\ &
= \fracc{\beta \theta}{\zeta + \nu t} (z^{-1}-1) \psi(z,t) + z^{-\beta}
.}{psi_dot_1}
In Appendix~\ref{app:pde} we solve this partial differential equation. Let us define
\all{
\begin{cases}
c \stackrel{\text{def}}{=} 1 - \left( \fracc {\zeta}{\zeta+\nu t}\right)^{\frac{\beta }{\nu}} \\ \\
F(z,t) \stackrel{\text{def}}{=} \fracc{ (\zeta+\nu t) }{\beta } \DD \sum_{m=0}^{\infty} (-1)^{m}\fracc{ z^{\frac{-\nu}{\beta }-m-\beta}}{\frac{\nu}{\beta } +m+\beta+\theta}
.
\end{cases}
}{defs}
Using these definitions, the solution to~\eqref{psi_dot_1} reads
\all{
& \psi(z,t) 
= F(z,t) \nonumber \\ &
+ z^{\theta}
\left(\fracc{z-c}{1-c}\right)^{-\theta}
\bigg[\psi \left( \fracc{z-c}{1-c} , 0\right) -F \left(\fracc{z-c}{1-c},0 \right) \bigg]
.}{psi_sol}
Note that $\psi \left( \frac{z-c}{1-c} , 0\right)$ is obtained by taking the Z-transform of the sequence $N_k(0)$ (which is given as the initial condition) and then replacing $z$ by $\frac{z-c}{1-c}$. In Appendix~\ref{app:inv} we take the inverse transform of  $\psi(z,t)$  to obtain $M_k(t)$. 
Taking the inverse transform of~\eqref{psi_sol}---which is given by~\eqref{Mk_FIN}---and dividing  the result
by the number of nodes at time $t$, we arrive at the following expression for the degree distribution of the network at time $t$:
\begin{widetext}


\all{
P_k(t)&
\resizebox{.93 \linewidth}{!}{$
= 
(1-c)^{\theta} c^k \fracc{N(0)}{N(0)+t}\DD \sum_{r=1} ^{k} P_r(0) \left( \fracc{1-c}{c} \right)^r \CC{k+\theta-1}{r+\theta-1}
+ \DD 
\fracc{ 1 }{ \beta } \fracc{ \zeta+ (2\beta +\theta) t }{N(0)+t}
\fracc{\Gamma(k+\theta) }{ \Gamma(\beta+\theta)} \fracc{\Gamma \left( \beta+ 2+\frac{\theta}{\beta} +\theta \right) }
{\Gamma \left( k+3+ +\frac{\theta}{\beta} + \theta \right) } u(k-\beta)
$}
\nonumber \\
&
- \fracc{ \zeta (1-c)^{\theta} c^k }{ \beta } 
\fracc{ \Gamma \left( \beta+ 2+\frac{\theta}{\beta} +\theta \right)}{N(0)+t}
\fracc{\Gamma(k+\theta) }{ \Gamma(\beta+\theta)}
\DD \sum_{r=\beta} ^{k}
\displaystyle \frac{ \left( \fracc{1-c}{c} \right)^r }{(k-r)! \Gamma \left( r+3+ \frac{\theta}{\beta} + \theta \right)},
}{Pk_ultimate}
where, as mentioned above, and repeated here for convenience of reference, we have 
$c= 1 - \left( \frac {\zeta}{\zeta+(2\beta+\theta) t}\right)^{\frac{\beta }{2\beta+\theta}} $, and also $\zeta= 2L(0)+\theta N(0) $ is obtained from initial conditions. Note that for the special case  of $\theta=0$, the result in~\eqref{Pk_ultimate} correctly reduces to that previously found in the literature~\cite{fotouhi2013network}.

\end{widetext}

The first term on the right hand side of~\eqref{Pk_ultimate} is the effect of initial nodes. In the long time limit, the $N(0)+t$ in the denominator makes this term vanish. Moreover, in the limit as $t \rightarrow \infty$, we have $c \rightarrow 1$, which means that every $(1-c)^r$ term as well as the $(1-c)^{\theta}$ prefactor tend to zero in the long time limit. Note that the $c^r$ in the denominator will not cause divergence, because the $c^k$ prefactor removes the singularity. So the first term on the right hand side of~\eqref{Pk_ultimate} vanishes in the long time limit, as we intuitively expect. 

The second term on the right hand side of~\eqref{Pk_ultimate} reaches a horizontal asymptote in the long time limit. In this limit, we have $ \frac{ \zeta+ (2\beta +\theta) t }{N(0)+t} \rightarrow 2\beta+\theta$. 

Finally, the last term on the right hand side of~\eqref{Pk_ultimate} vanishes in the long time limit for the same reasons delineated above for the first term. So in the steady state, the first and third terms have no share in the degree distribution, and the second term dominates. We have: 

\all{
\resizebox{\linewidth}{!}{$
\lim_{t\rightarrow \infty} P_k(t)= 
\DD 
\left( 2+\fracc{\theta}{\beta}\right)
\fracc{\Gamma(k+\theta) }{ \Gamma(\beta+\theta)} \fracc{\Gamma \left( \beta+ 2+\frac{\theta}{\beta} +\theta \right) }
{\Gamma \left( k+3+\frac{\theta}{\beta} + \theta \right) } u(k-\beta)
.
$}
}{Pk_ss}

%
%

This is in agreement with the results in~\cite{dorogovtsev2000structure}. 
Finally, we can set $\theta=0$ to recover the degree distribution of the conventional Barab{\'a}si-Albert model: 

\all{
\lim_{t\rightarrow \infty} P_k^{\textnormal{BA}}(t) &= 
\DD 
2
\fracc{(k-1)! }{(\beta-1)!} \fracc{(\beta+1)! }
{(k+2)! } u(k-\beta) 
\nonumber \\ &
= \fracc{2 \beta(\beta+1)}{k(k+1)(k+2)} u(k-\beta)
.}{Pk_ss}

This is in agreement with the long-known result, as given for example in~\cite{dorogovtsev2000structure,krapivsky2001organization}.

\begin{figure}[h]
\centering
\includegraphics[width=.95 \columnwidth]{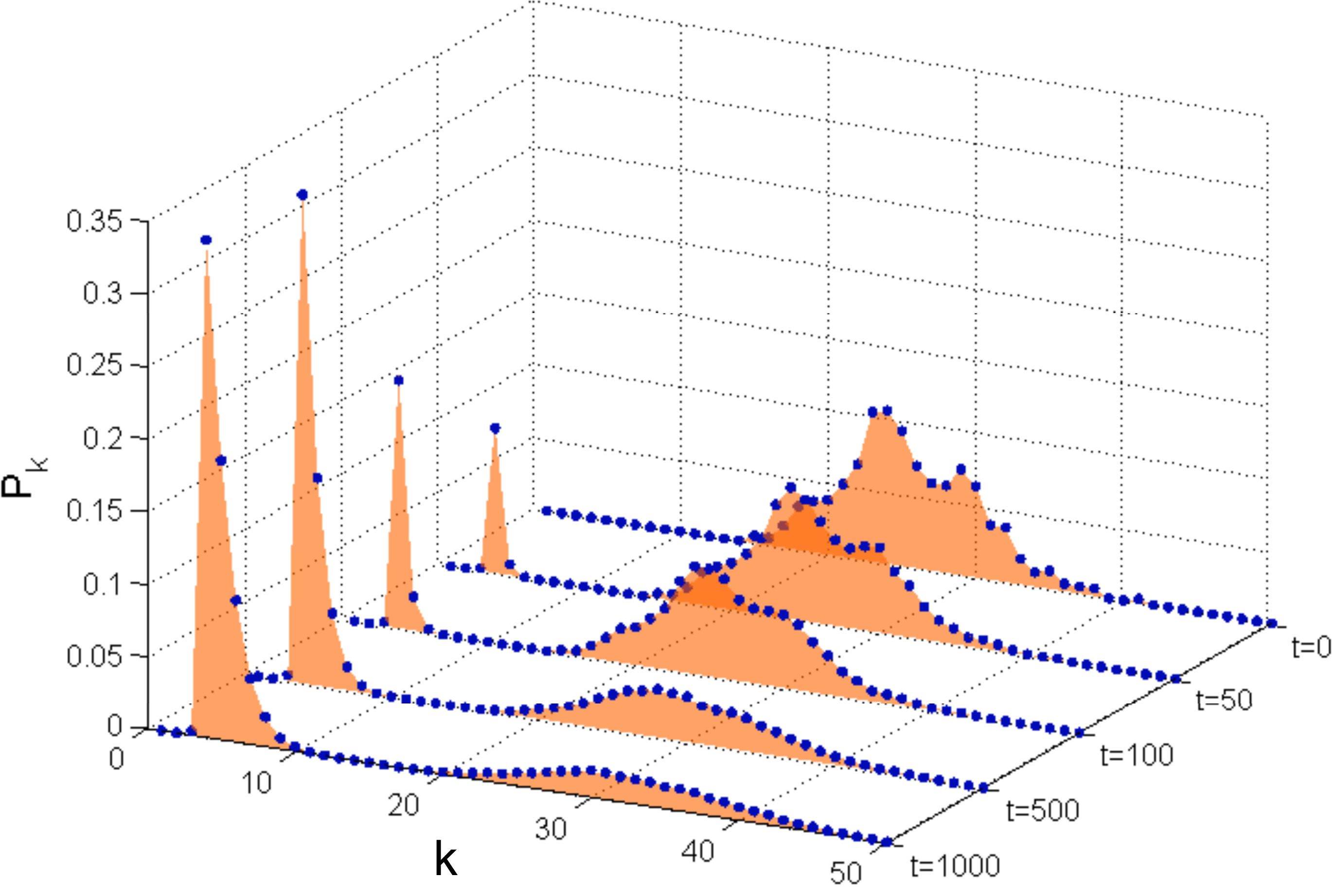}
\caption[Figure ]%
{ (Color online) Simulation results for the assessment of the theoretical prediction of the time-dependent degree distribution~\eqref{Pk_ultimate}. Markers represent simulation results, averaged over 50 Monte Carlo trials. Shaded areas pertain to the theoretical prediction. Shades are used instead of linear plots for better visibility. 
The initial substrate is a $SW(3,400,0.05)$ network, as described in the text. The degree distribution of the initial substrate corresponds to the curve depicted for $t=0$. The parameters of the growth mechanism are $\beta=4$ and $\theta=5.2$, which are example values chosen for expository purposes. As time progresses, the initial lump, which is concentrated around the mean degree of the initial substrate---that can be obtained from ${2\times 3 + 0.05 \times (400-2 \times 3)}$, which is close to 26---loses its mass, because new nodes are being introduced to the network. Newcomers have initial degree $\beta=4$, hence probability mass shifts leftwards. Note that the time axis is not linear: few timesteps are chosen for illustrative purposes to maintain convenience of vision. }
\label{ZPK_SW_p_5sadom_N0_400_d_3_beta_4_th_5}
\end{figure}

To assess the accuracy of the theoretical prediction given in~\eqref{Pk_ultimate}, we run Monte-Carlo simulations to compare the theoretical prediction with simulation results. For the first setup, we consider a small-world network, constructed as follows. Consider a $2b$-regular ring of $N(0)$ nodes, which is similar to a ring, but instead of each node being connected to one node to the left and one node to the right, it is connected to $b$ nodes from each side. Then, create each nonexistent link with probability $p_{SW}$. Hereinafter, we denote a network that is constructed this way by $SW(b,N(0),p_{SW})$, where SW stands for small world. The first simulation comprises a $SW(3,400,0.05)$ network. For the growth process, example values of $\beta=4$ and $\theta=5.2$ are considered. Figure~\ref{ZPK_SW_p_5sadom_N0_400_d_3_beta_4_th_5} depicts the degree distribution at several arbitrary timesteps, to emphasize the strength of the analysis presented here: that the analysis is in no way limited to the long-time limit. The remarkable accuracy of the prediction is visible. Note that the evolution of the degree distribution can be grasped conceptually as follows. The initial substrate has a concentrated degree distribution, which is expected from   SW networks. As the network grows, nodes with initial degree $\beta$ are introduced to the network. Hence, a peak emerges at $k=\beta$, more probability mass moves from the initial peak towards the new peak at $\beta$ as time progresses, since more and more nodes with degree $\beta$ are being introduced. The initial lump in the degree distribution---which was the effect of the nodes in the initial substrate---becomes smaller as it loses mass throughout time, and lends its mass to the new nodes, and in the limit as $t \rightarrow \infty$, the effect of initial conditions would vanish altgether and a power-law degree distribution would materialize.

\begin{figure}[h]
\centering
\includegraphics[width=.95 \columnwidth]{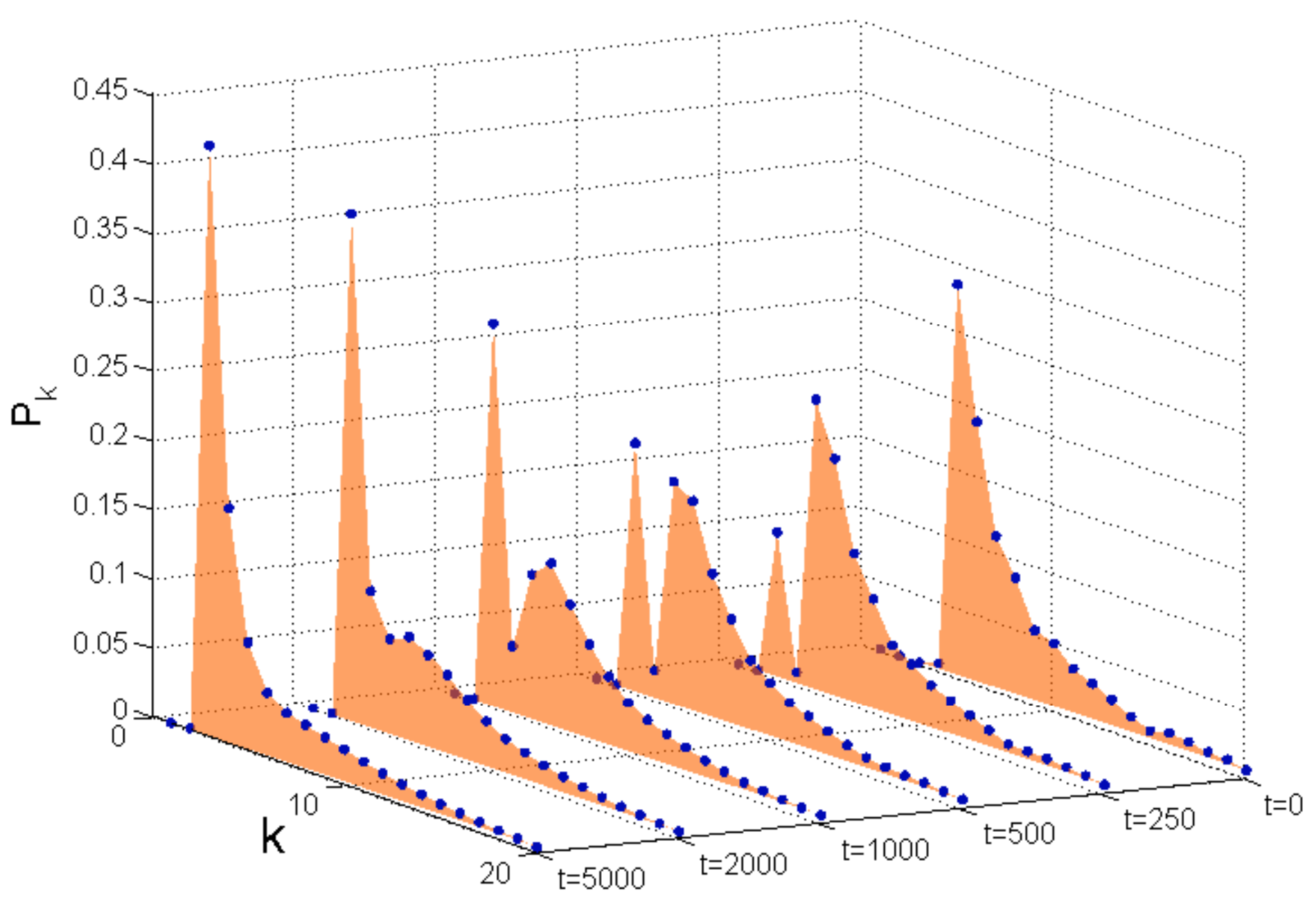}
\caption[Figure ]%
{ (Color online) Simulation results for the assessment of the theoretical prediction of the time-dependent degree distribution~\eqref{Pk_ultimate} (The view angle is rotated to achieve better visibility, hence the right-to-left ordering of timesteps.). The markers represent simulation results, and filled areas represent theoretical predictions. The simulation results are averaged over 50 Monte Carlo trials. The initial substrate is a BA network of 2000 nodes, grown with $\beta=5$. It is visible that at $t=0$ the peak of the degree distribution is at $k=\beta_0=5$. The degree distribution of the initial substrate is approximately power-law, starting from $k=5$. As new nodes are introduced, the population of nodes with degree $k=\beta=3$ increases. The lump concentrated around $k=5$ loses its mass as time progresses, and the probability mass moves towards $k=3$. In the limit as $t \rightarrow \infty$, we get a power-law degree distribution starting at $k=3$. 
}
\label{ZPK_N0_2000_beta0_5_beta_3_BA}
\end{figure}

As the second simulation setting for the assessment of the theoretical prediction on the degree-distribution, we consider the following setup. First, consider a complete graph of $\beta_0$ nodes, and let the network grow according to the shifted-linear model discussed in this paper until its size reaches some given $N(0)>\beta$. We denote such a network by $SL(\beta_0,N(0),\theta)$. For the second simulation, we take a $SL(5 , 2000 , 0 )$ network, and apply to it a growth process with different $\beta$ and $\theta$ than those with which the initial network was constructed. We chose the example values $\beta= 3 $ and $\theta= 0 $. 
Simulation results are depicted along with theoretical prediction~\eqref{Pk_ultimate} in Figure~\ref{ZPK_N0_2000_beta0_5_beta_3_BA}. We chose $\theta=0$ for convenience of visible interpretation: in the substrate, we have a classical BA model, with the degree distribution peak in at $\beta_0$.
As the network grows, the peak of the probability distribution moves from $\beta_0$ towards $\beta$, due to the introduction of new nodes who all have degree $\beta$  upon birth, which results in probability mass being transferred towards $\beta$. The degree distribution at intermediate times are all captured by the theoretical prediction$\eqref{Pk_ultimate}$ with remarkable accuracy.

\begin{figure}[h]
\centering
\includegraphics[width=.95 \columnwidth]{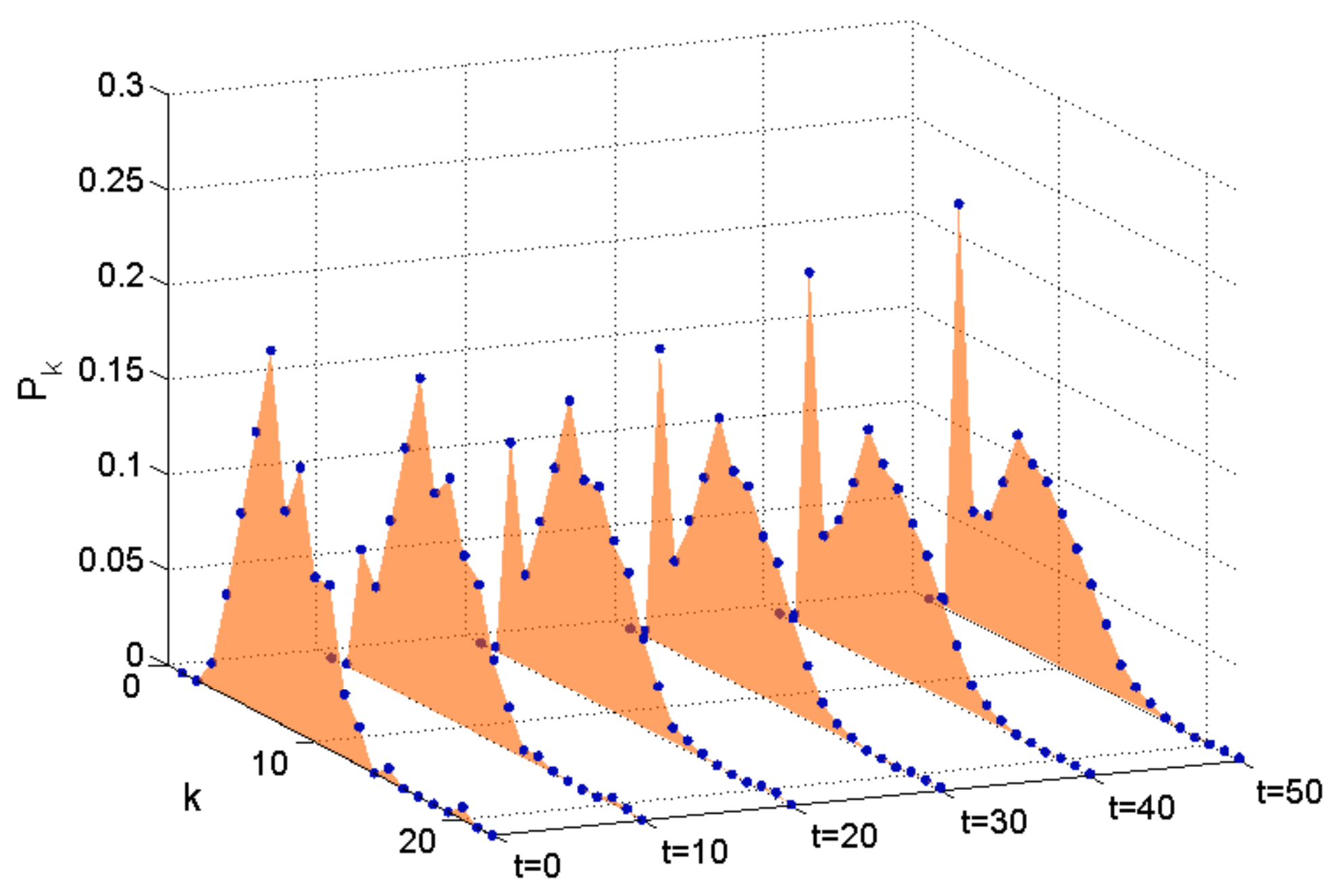}
\caption[Figure ]%
{ (Color online) Simulation results for the assessment of the theoretical prediction of the time-dependent degree distribution~\eqref{Pk_ultimate}. The substrate is an ER  network with 150 nodes. The existence probability of each link is 0.05. The growth parameters are $\beta=3$ and $\theta=1.9$. 
The markers represent simulation results, and filled areas represent theoretical predictions. The simulation results are averaged over 100 Monte Carlo trials.  
}
\label{Z_Pk_khoshgel_ER_5sadom_N0_150_beta_3}
\end{figure}

As the third example, we study the accuracy of the theoretical prediction for the  Erd\H{o}s-R\'enyi (hereinafter ER) graph. We consider an ER network of 150 nodes, where the probability of existence for each link is 0.05. The growth parameters are $\theta=2.1$ and $\beta=3$. The results are depicted in Figure~\ref{Z_Pk_khoshgel_ER_5sadom_N0_150_beta_3}. 

We also   test the theoretical prediction of $P_k(t)$ on real networks. In Appendix~\ref{app:Pk_t}, we use the social network of dolphins~\cite{lusseau2003bottlenose}, the network of collaborations between scholars working in the field of network science~\cite{newman2006finding}, as well as condensed matter~\cite{newman2001structure}  as substrates, in order  to simulate the growth mechanism.  Theoretical predictions and simulation results consistently agree.

\section{The SIS model}
A  basic  compartmental model of the spread of infectious diseases is the SIS model~\cite{RevModPhys.87.925}, in which each node  is  either susceptible (S) and can get infected, or  it is   infected (I) and can recover and become susceptible again. The transmission rate of the disease per infected contact is denoted by $\eta$,  and the recovery rate is denoted by $\mu$. The control parameter ${\lambda \stackrel{\text{def}}{=} \frac{\eta}{\mu}}$ is used to characterize the system. 
We consider the degree-based mean-field approximation~\cite{boguna2003absence,pastor2002epidemic}, which is  an analytically-parsimonious approach for incorporating degree heterogeneity\cite{RevModPhys.87.925}. It is based on an uncorrelated approximation,  and is also  able of providing surprisingly accurate results for  some correlated networks~\cite{gleeson2012accuracy,melnik2011unreasonable}. The epidemic threshold  under this approximation is given by 
\all{
\Lambda_{\textnormal{SIS}}=\fracc{\ave{k}}{\ave{k^2}}
.}{th_sis}
Thus, the theoretical prediction for the epidemic threshold at time $t$ is given by
\all{
\Lambda_{\textnormal{SIS}} (t) = \fracc{\sum_{k} k P_k(t) }{\sum_{k} k^2 P_k(t)}
.}{th_sis_t}

Thus, we can utilize the expression for $P_k(t)$ given in~\eqref{Pk_ultimate} in order to predict the epidemic threshold at time $t$. We  examine the accuracy of this prediction on several example topologies. 

\begin{figure}[h]
\centering
\includegraphics[width=.95 \columnwidth]{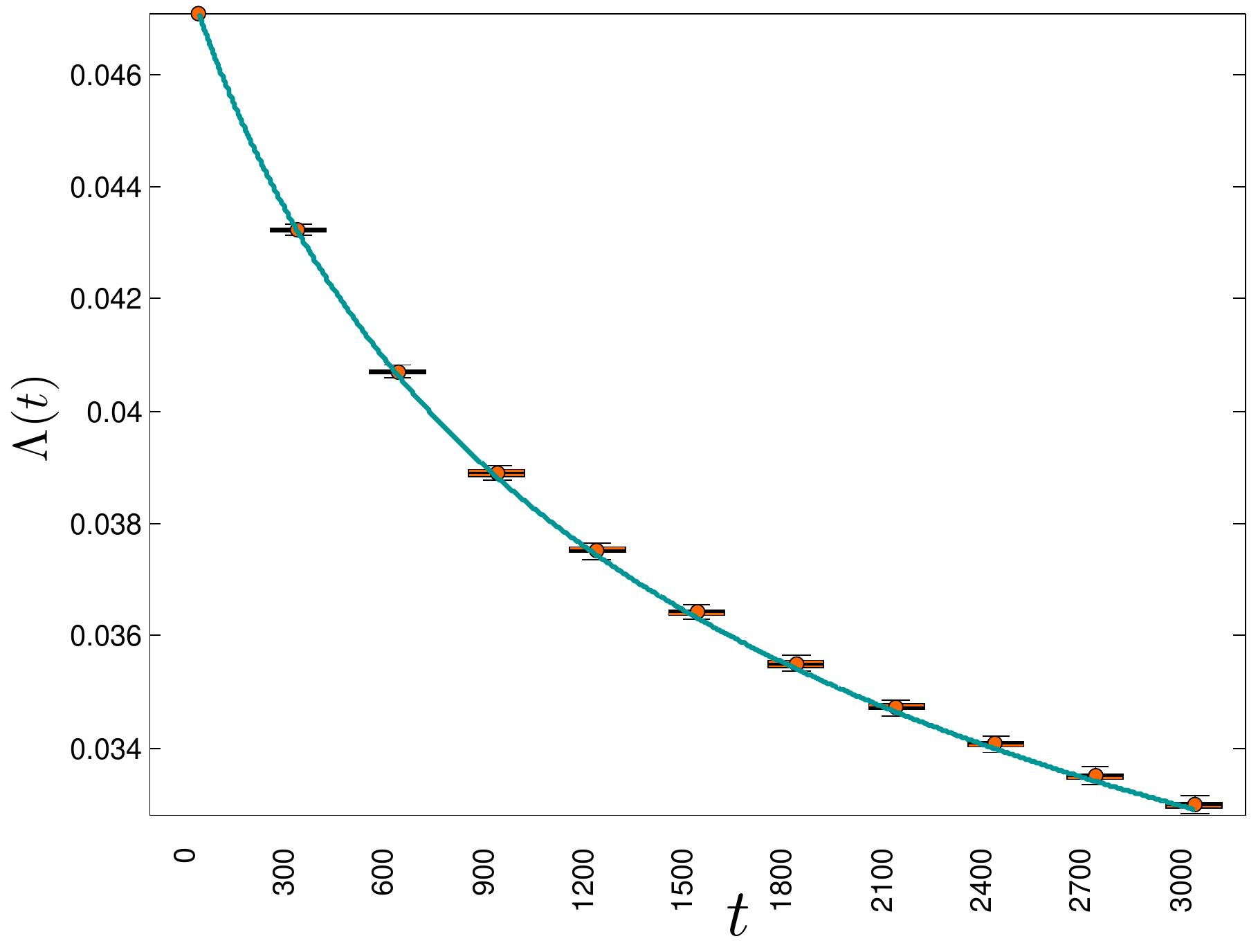}
\caption[Figure ]%
{ (Color online) Temporal evolution of the epidemic threshold for the SIS model. The initial substrate is an Erd\H{o}s-R\'enyi network of 1000 nodes, where the probability of existence for each link is equal to 0.02. 
The growth parameters are $\beta=12$ and $\theta=3.2$. 
The error bars pertain to 1000 Monte Carlo simulations. The solid line represents the theoretical prediction~\eqref{th_sis_t}.
The inset shows  the theoretical prediction for the second moment of the degree distribution,  that  is invoked in the calculation of the epidemic threshold~\eqref{th_sis}, which also exhibits good agreement between theoretical prediction  and simulation results. 
}
\label{Z_ER_p_2sadom_N0_1000_T_3000_beta_12_th_3}
\end{figure}

Consider an ER graph of 1000 nodes, with probability of the existence of each link being equal to 0.02. The epidemic threshold as given by~\eqref{th_sis} is applicable to this setting, due to lack of degree-degree correlations. The degree distribution is straightforward to characterize; it is binomial with parameters 0.02 and 999, thus the epidemic threshold is  equal to  
0.47. What happens to the epidemic threshold if the network begins to grow and new  nodes  are introduced to the system? Figure~\ref{Z_ER_p_2sadom_N0_1000_T_3000_beta_12_th_3} pertains to the setup where $\beta=12$ and $\theta=3.2$. It can be observed that the epidemic threshold diminishes as the network grows. This means that as new nodes enter the network, they  decrease the resilience of the network against epidemic outbreaks. Note that the simulations only pertain to the network growth process, and the epidemic thresholds pertain to potential outbreaks, and not the concurrent evolution of network size and disease spread. This is in accordance with the adiabatic assumption discussed above.

\begin{figure}[h]
\centering
\includegraphics[width=.95 \columnwidth]{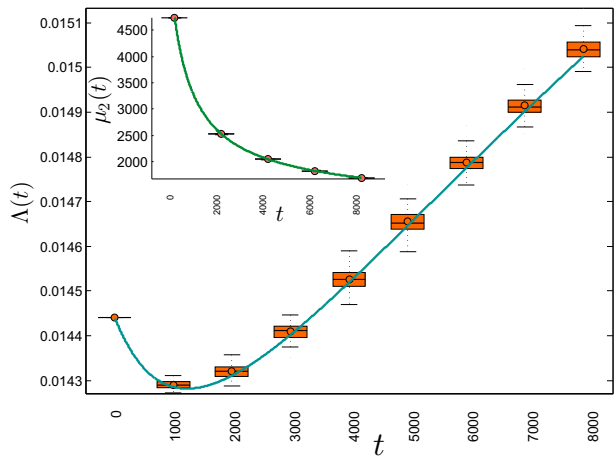}
\caption[Figure ]%
{ (Color online) Temporal evolution of the epidemic threshold for the SIS model. The initial substrate is a small-world network, namely, a $SW(10,1000,0.05)$ network. 
The growth parameters are $\beta=10$ and $\theta=12.5$. 
The error bars pertain to 1000 Monte Carlo simulations. The solid line represents the theoretical prediction~\eqref{th_sis_t}.
The inset shows  the theoretical prediction for the second moment of the degree distribution.
}
\label{Z_SW_p_5sadom_N0_1000_T_8000_beta_10_th_10_d_10}
\end{figure}

The second   example we consider  is the SW network. Consider a $SW(10,1000,0.05)$. The degrees of adjacent nodes are uncorrelated by construction, since attachments are agnostic on  destination degrees. Figure~\ref{Z_SW_p_5sadom_N0_1000_T_8000_beta_10_th_10_d_10} depicts the temporal evolution of the epidemic threshold. The growth parameters are $\beta=10$ and $\theta=12.5$ (we choose noninteger example values for $\theta$ merely to reflect the fact that $\theta$ is arbitrary, and need not be an integer). 
The simulation results are again in good agreement with the theoretical prediction. As Figure~\ref{Z_SW_p_5sadom_N0_1000_T_8000_beta_10_th_10_d_10}  illustrates,  the epidemic threshold first decreases up to some time, and increases afterwards. This means that after a short initial period, the incoming nodes begin to increase the resilience of the system against epidemic outbreaks.

\begin{figure}[h]
\centering
\includegraphics[width=.95 \columnwidth]{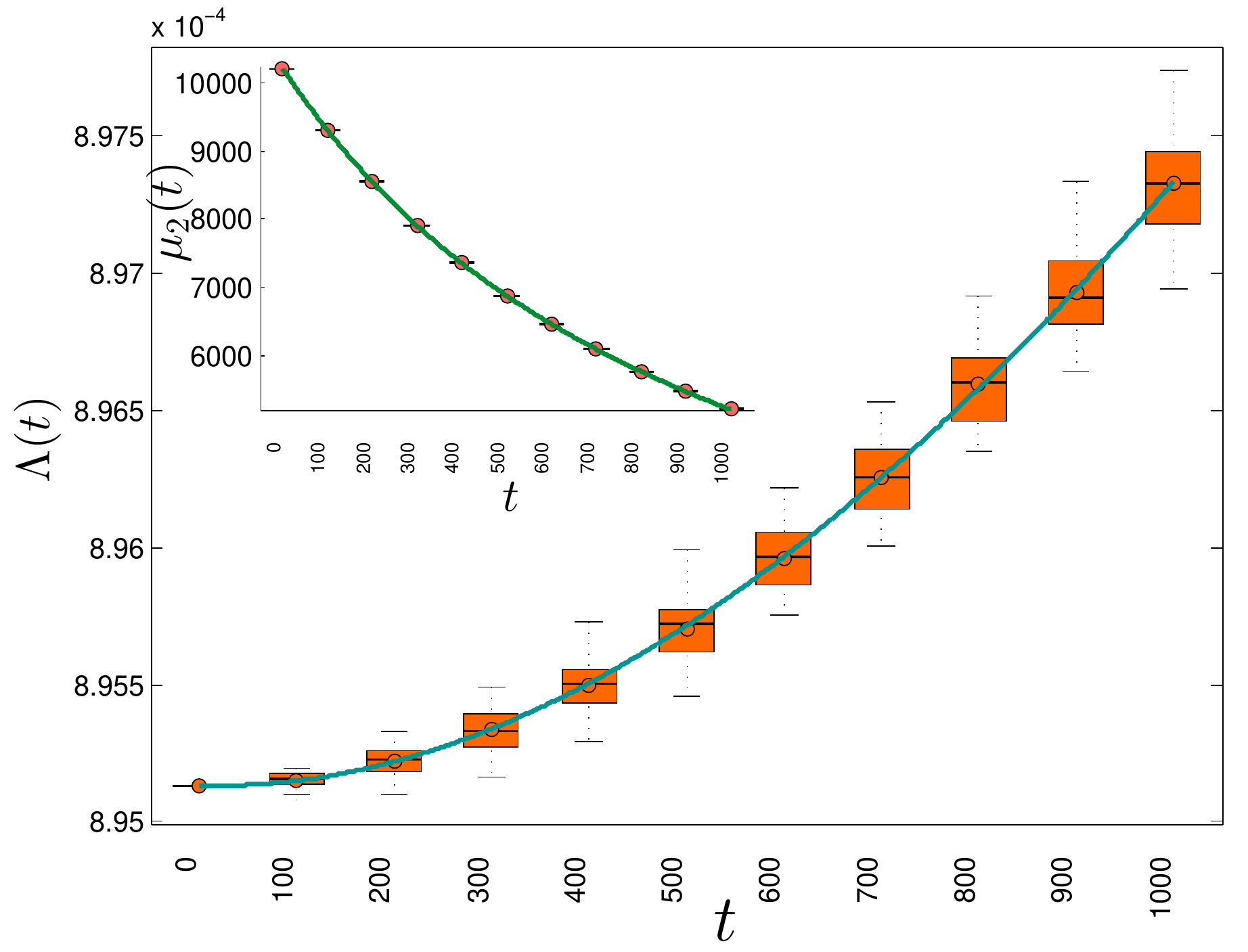}
\caption[Figure ]%
{ (Color online) Temporal evolution of the epidemic threshold for the SIS model. The initial substrate is a an ER network with 1000 nodes. the existence probability of  links is 0.1. 
The growth parameters are $\beta=1$ and $\theta=50$. 
The error bars pertain to 1000 Monte Carlo simulations. The solid line represents the theoretical prediction~\eqref{th_sis_t}.
The inset shows  the theoretical prediction for the second moment of the degree distribution.
As a result of the addition of new incoming nodes to the network, the epidemic threshold increases, making the network more  resilient to endemic outbreaks. 
}
\label{ZZZ_afzayeshi_SIS_ER_theta_50_beta_1_N0_1000}
\end{figure}

We can also have a setting in which the epidemic threshold begins to grow monotonically as the incoming nodes enter. Consider an ER network of 1000 nodes and let the existence probability of each link be 0.1.  We consider a growth mechanism on this substrate. The results for the growth parameters $\beta=1$ and $\theta=50$ are depicted in Figure~\ref{ZZZ_afzayeshi_SIS_ER_theta_50_beta_1_N0_1000}. The epidemic threshold grows monotonically, which indicates that incoming nodes bring more resilience to the network against endemic outbreaks, without a transient phase---which was the case in Figure~\ref{Z_SW_p_5sadom_N0_1000_T_8000_beta_10_th_10_d_10}.

To verify the accuracy of the theoretical predictions for diverse substrates, in addition to the ones considered above, we have considered more synthetic networks with different topologies  (namely, a ring, a complete graph, and a random recursive tree~\cite{dorogovtsev2008transition}). The results of these topologies are  presented in Appendix~\ref{app:more_SIS}---despite lending more credibility to the theoretical predictions, they offer no new conceptual insight, so we omit them from the main text.

\begin{figure}[h]
\centering
\includegraphics[width=.95 \columnwidth]{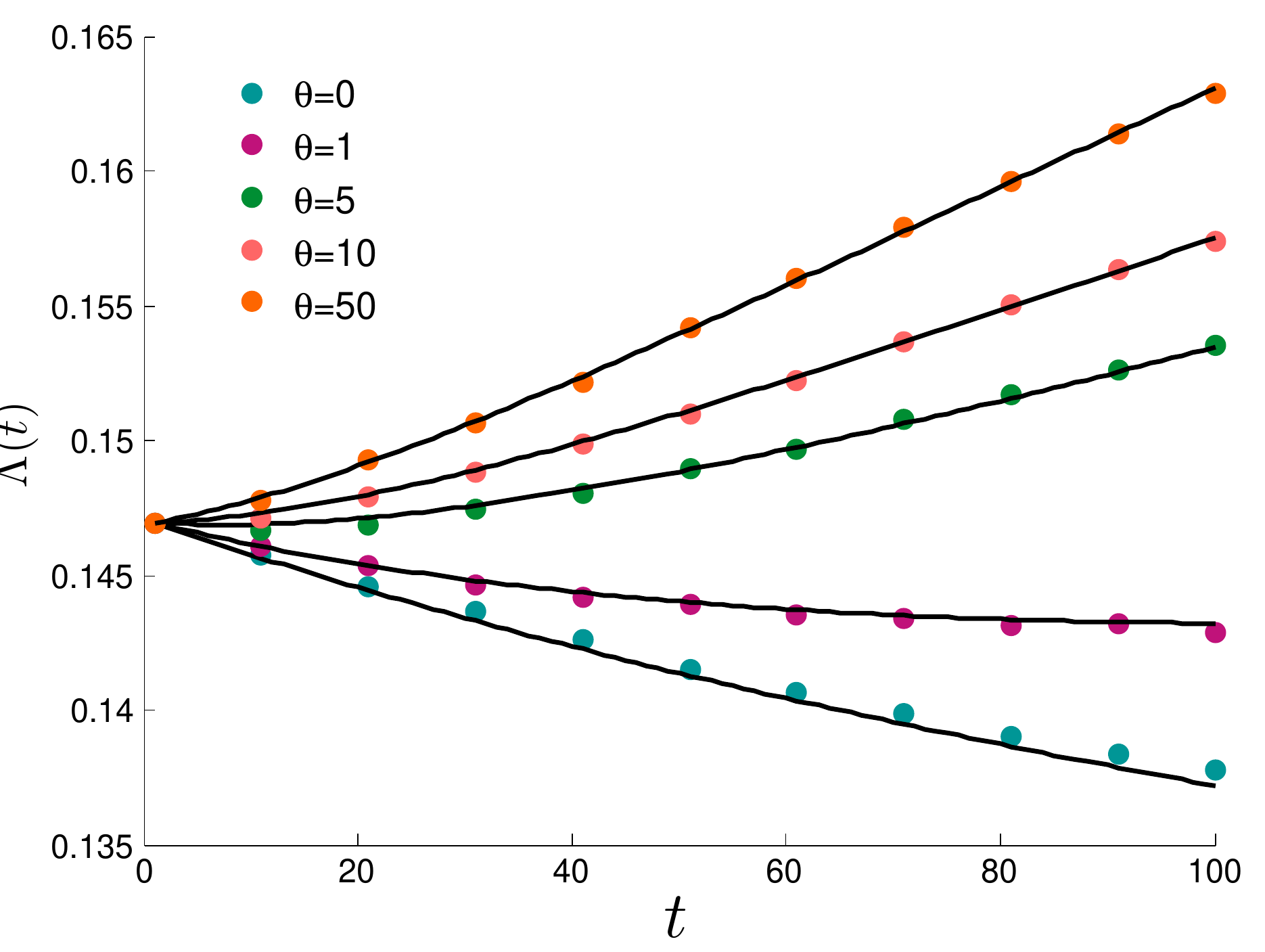}
\caption[Figure ]%
{ (Color online) Temporal evolution of the epidemic threshold for the SIS model. The initial substrate is the social network of dolphins~\cite{lusseau2003bottlenose}.  The initial connectivity of incoming nodes is $\beta=1$. 
The markers represent simulation results, averaged over 100 Monte Carlo trials. The solid line represents the theoretical prediction~\eqref{th_sis_t}.
}
\label{Z_dolphin_SIS_beta_1}
\end{figure}

Finally, let us consider an example where the substrate is a real network. We consider  the social network of dolphins~\cite{lusseau2003bottlenose}. It has 62 nodes. We consider the initial connectivity of $\beta=1$, and investigate the temporal evolution of the epidemic threshold for different values of $\theta$. The results are depicted in Figure~\ref{Z_dolphin_SIS_beta_1}. The graph indicates that as $\theta$ increases, the future epidemic threshold will  increase at a faster rate. Conversely, for low values of $\theta$ (e.g., for $\theta=0$, which is the conventional preferential attachment in the BA model), the epidemic threshold decreases over time, which means that the introduction of new nodes  makes the system more susceptible to future endemic outbreaks.

\section{The SIR model}
\textbf{Epidemic Threshold:} 
Now we consider the SIR model of epidemic spreading, as analyzed in~\cite{boguna2003epidemic,lindquist2011effective}. Unlike the $SIS$ model, the infected individuals do not return to the S state. Instead, they either recover and remain immune thereafter, or they die and get removed. The epidemic threshold is given by 
\all{
\Lambda_{\textnormal{SIR}}=\fracc{\ave{k}}{\ave{k^2} - \ave{k}}
.}{th_sir}
Thus, the theoretical prediction for the epidemic threshold at time $t$ is given by
\all{
\Lambda_{\textnormal{SIR}} (t) = \fracc{\sum_{k} k P_k(t) }{\sum_{k} (k^2-k) P_k(t)}
.}{th_sir_t}

\begin{figure}[h]
\centering
\includegraphics[width=.95 \columnwidth]{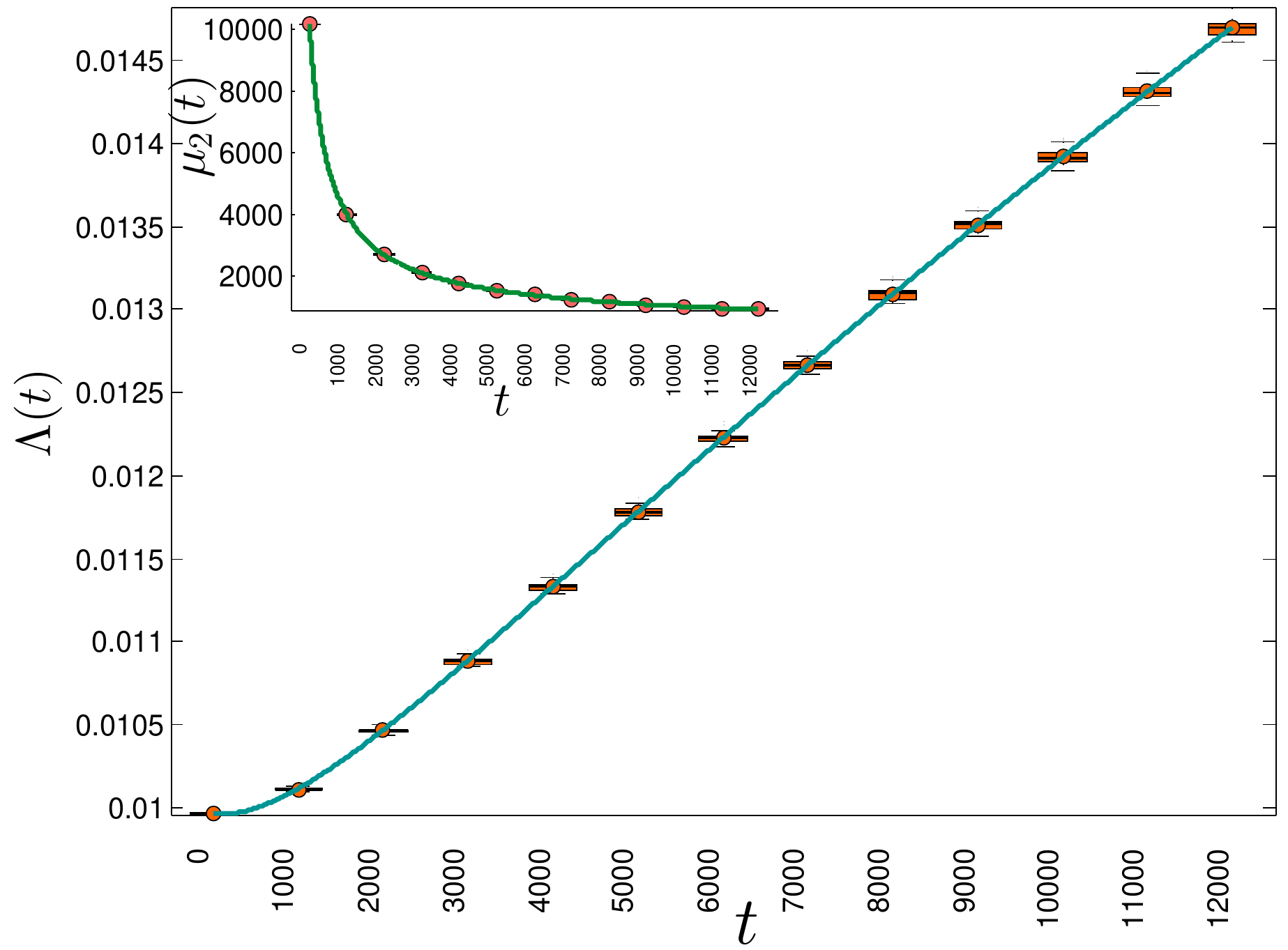}
\caption[Figure ]%
{ (Color online) Temporal evolution of the epidemic threshold for the SIR model. The initial substrate is an Erd\H{o}s-R\'enyi network with 500 nodes, with link creation probability of 0.2. 
The growth parameters are $\beta=5$ and $\theta=20$. 
The error bars pertain to 1000 Monte Carlo simulations. The solid line represents the theoretical prediction~\eqref{th_sir_t}.
The inset shows  the theoretical prediction for the second moment of the degree distribution, which is invoked in the calculation of the epidemic threshold~\eqref{th_sir_t}.
}
\label{ZZZZZ_ER_p_2dahom_N0_500_T_12000_beta_5_th_20_khoob}
\end{figure}

To test this result, we first consider an ER graph of 500 nodes, with link creation probability 0.2. According to~\eqref{th_sir}, the epidemic threshold of this graph is  
0.01 Now let us consider a growth mechanism with initial attractiveness $\theta=20$ and initial connectivity $\beta=5$.
The results are presented in Figure~\ref{ZZZZZ_ER_p_2dahom_N0_500_T_12000_beta_5_th_20_khoob}. 
When the incoming nodes are added to the network, the epidemic threshold increases.

Now let us shift our focus and study the evolution of a quantity  that  is  also  central in the mathematical models of epidemic disease, which is the basic reproduction number (denoted by $\mathcal{R}_0$). It is the expected number of susceptible individuals that an  infected node transmits the disease to,
 before recovery, in a fully-susceptible population. If $\mathcal{R}_0<1$, the disease will die out. It can  cause an outbreak otherwise. We only study $\mathcal{R}_0$ for the SIR model for space limitations (the case of SIS is conceptually congruent to SIR).

\textbf{Basic Reproduction Number:}
We now use the theoretical findings to make predictions about the temporal behavior of the basic reproduction number, which is indicative of the potential of a certain disease to become endemic. For the SIR model, as studied in~\cite{newman2002spread,lindquist2011effective}, the basic reproduction number of a disease with transmission rate $\eta$ and recovery rate $\mu$ under the mean-field approximation is given by
\all{
\mathcal{R}_0= \fracc{\eta}{\eta+\mu} \left( \ave{k} - 1 + \fracc{\ave{k^2}-\ave{k}^2}{\ave{k}} \right) 
.}{R0_SIR}

Thus, at arbitrary time $t$, we have: 
\all{
&\mathcal{R}_0(t) = 
\nonumber \\ &
\resizebox{.95\linewidth}{!}{$
\fracc{\eta}{\eta+\mu} \left[ \sum_k k P_k(t) - 1 + \fracc{ \sum_k k^2 P_k(t)- \left( \sum_k k P_k(t)\right)^2 }{ \sum_k k P_k(t)} \right]
.
$}
}{R0_SIR_t}

\begin{figure}[h]
\centering
\includegraphics[width=.9 \columnwidth]{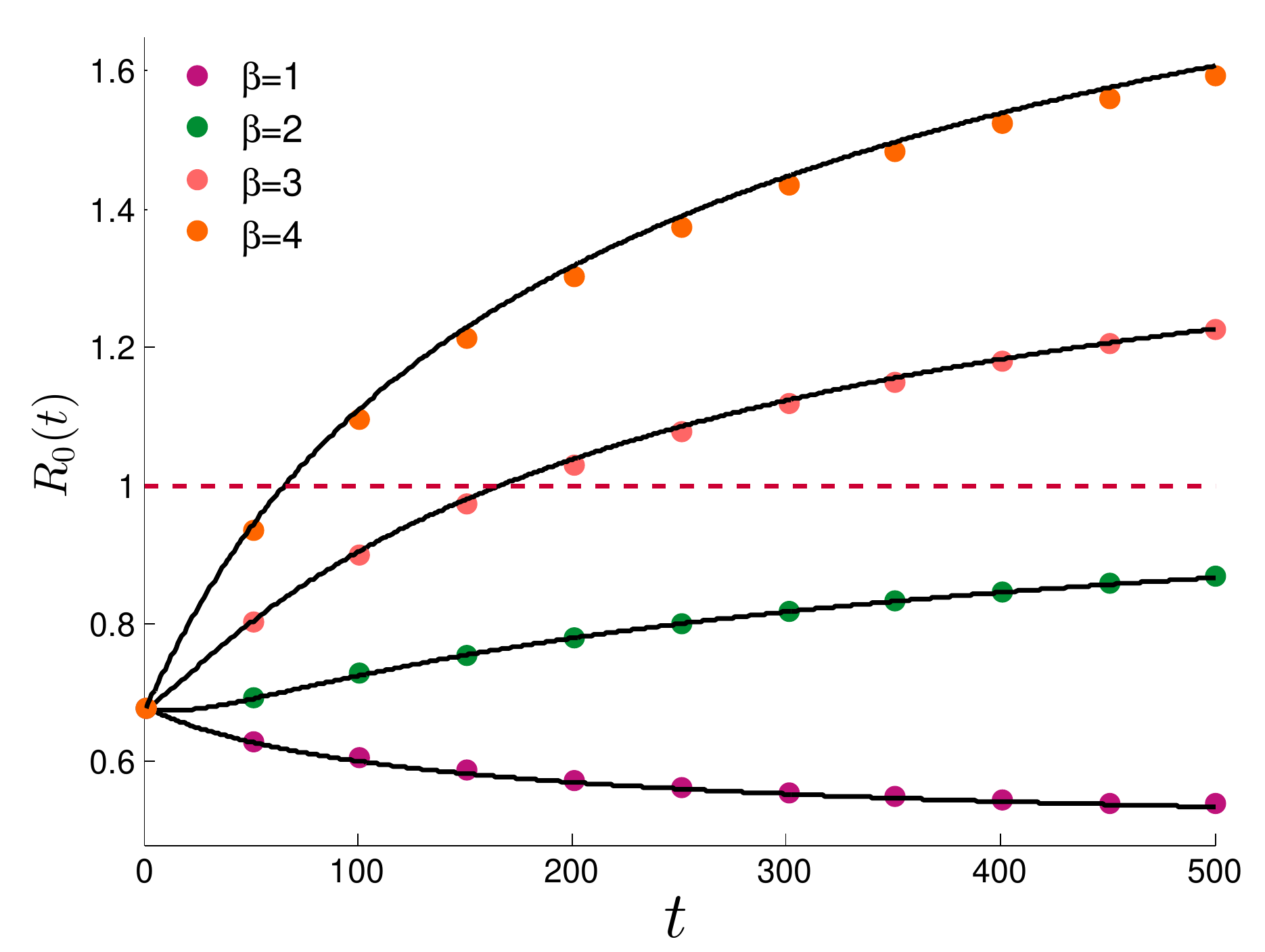}
\caption[Figure ]%
{ (Color online) Temporal evolution of the basic reproduction number in the SIR model. The initial substrate is a BA tree with 100 nodes. 
The initial attractiveness is $\theta=5.2$. The transmission rate is $\eta=0.05$ and the recovery rate is $\mu=0.2$.
The solid markers are averaged over 100 Monte Carlo trials. The solid line represents the theoretical prediction~\eqref{R0_SIR_t}. 
The initial network is secure, as $\mathcal{R}_0$ is below unity. As the network grows, the incoming node can drive $\mathcal{R}_0$ above unity, making the system susceptible to endemic disease outbreaks. 
}
\label{Z_R0_SIR_N0_100_beta0_1_th_5_beta_1_ta_4}
\end{figure}

Let us investigate the accuracy of this result. 
For simulation purposes, we first consider a tree, namely, a BA graph of 100 nodes with $\beta=1$. That is, the growth process begins with a single node and each incoming node attaches to one existing node selected according to degree-proportional probabilities. Figure~\ref{Z_R0_SIR_N0_100_beta0_1_th_5_beta_1_ta_4} depicts the simulation results and theoretical predictions for example values of $\nu=0.05$ and $\mu=0.2$, and the initial attractiveness is $\theta=5.2$. It can be observed from Figure~\ref{Z_R0_SIR_N0_100_beta0_1_th_5_beta_1_ta_4} that, although $\mathcal{R}_0$ for the initial substrate is below unity, incoming nodes can elevate the value of $\mathcal{R}_0$ and drive the network towards  an endemic potential. The higher the value of $\beta$ is, the more $\mathcal{R}_0$ will increase in time. This means that if the incoming nodes are highly connected, they can be detrimental to the epidemic properties of the system, which is intuitively expected. 

\begin{figure}[h]
\centering
\includegraphics[width=.9 \columnwidth]{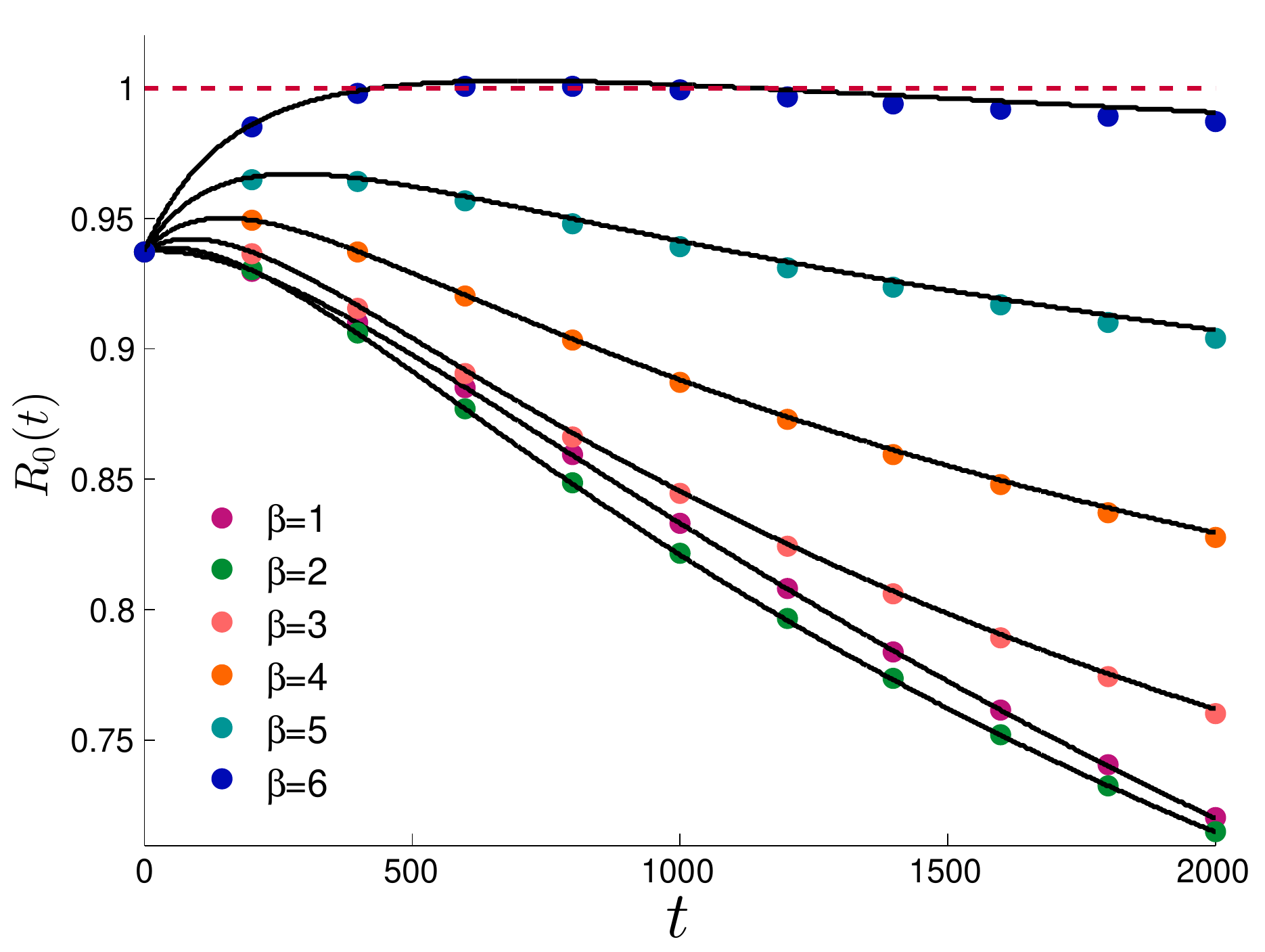}
\caption[Figure ]%
{ (Color online) Temporal evolution of the basic reproduction number in the SIR model. The initial substrate is an  ER  network with 200 nodes. The probability of existence of links is 0.1. 
The initial attractiveness in the growth process is $\theta=10$. The transmission rate is $\eta=0.01$ and the recovery rate is $\mu=0.2$.
The solid markers are averaged over 100 Monte Carlo trials. The solid line represents the theoretical prediction~\eqref{R0_SIR_t}. 
The initial network is secure, as $\mathcal{R}_0$ is below unity. As the network grows, if $\beta$ is sufficiently large, incoming node can drive $\mathcal{R}_0$ above unity, making the system susceptible to endemic disease outbreaks. 
}
\label{Z_R0_SIR_N0_200_ER_p_1dahom_th_10}
\end{figure}

For the second case study, we consider an uncorrelated network, namely, an ER network with 200 nodes, with the probability of existence for links equal to 0.1. The initial attractiveness is $\theta=10$. The transmission rate is $\eta=0.01$ and the recovery rate is $\mu=0.2$. The results are depicted in Figure~\ref{Z_R0_SIR_N0_200_ER_p_1dahom_th_10}. Similar to the case of the BA tree, for large enough values of $\beta$, the incoming nodes are able to increase the basic reproduction number above unity.

\begin{figure}[h]
\centering
\includegraphics[width=.9 \columnwidth]{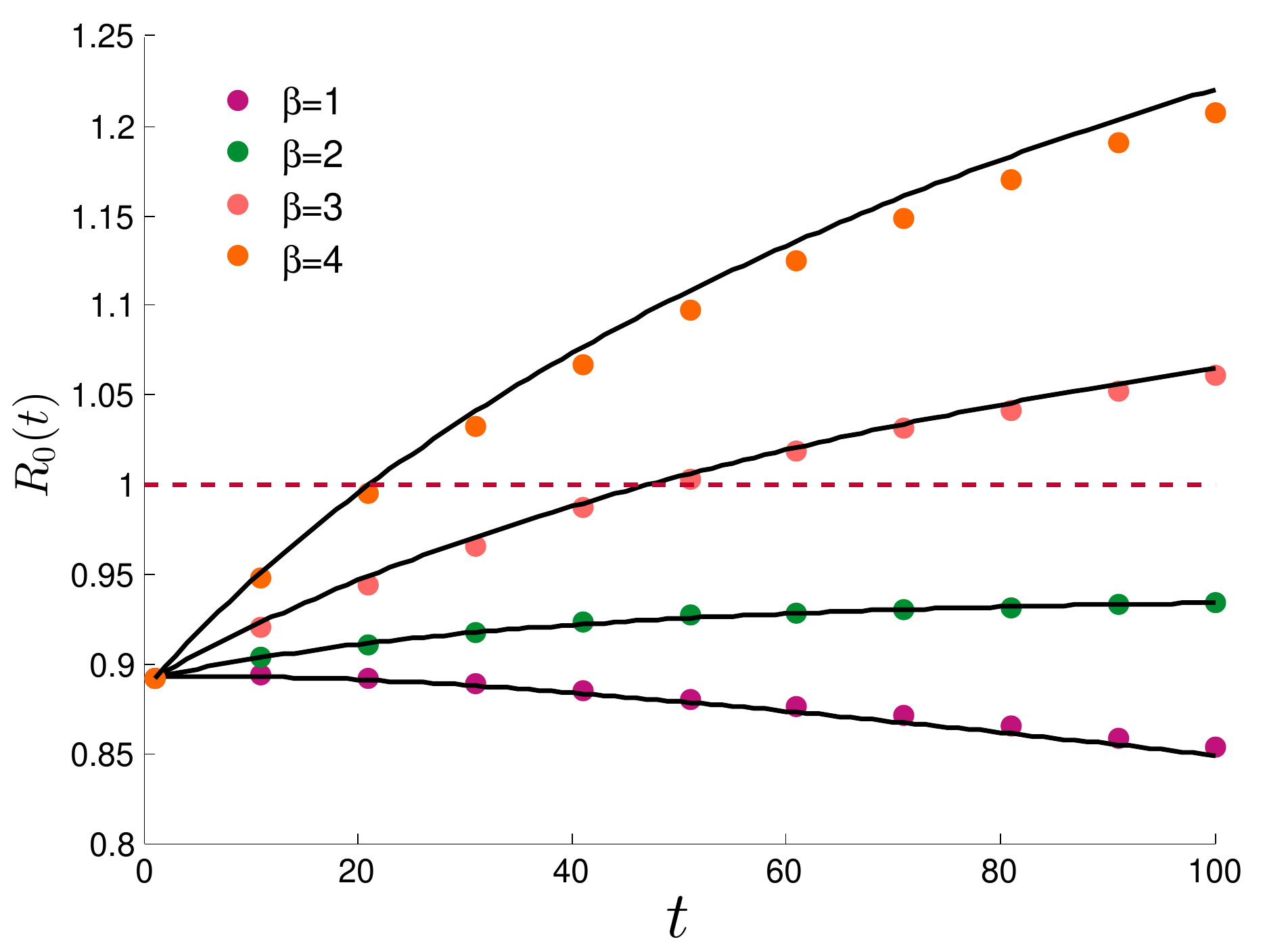}
\caption[Figure ]%
{ (Color online) Temporal evolution of the basic reproduction number in the SIR model for the social network of dolphins. 
The initial attractiveness in the growth process is $\theta=5$. The transmission rate is $\eta=0.04$ and the recovery rate is $\mu=0.22$.
The solid markers are averaged over 100 Monte Carlo trials. The solid line represents the theoretical prediction~\eqref{R0_SIR_t}. 
It  is readily discernable that higher values of $\beta$  result in   faster rates of increase in $\mathcal{R}_0$. This means that the more gregarious the newcomers are, the more the network will be driven towards a potential endemic. 
}
\label{Z_dolphin_R0_beta_avaz_th_5}
\end{figure}

\begin{figure}[h]
\centering
\includegraphics[width=.9 \columnwidth]{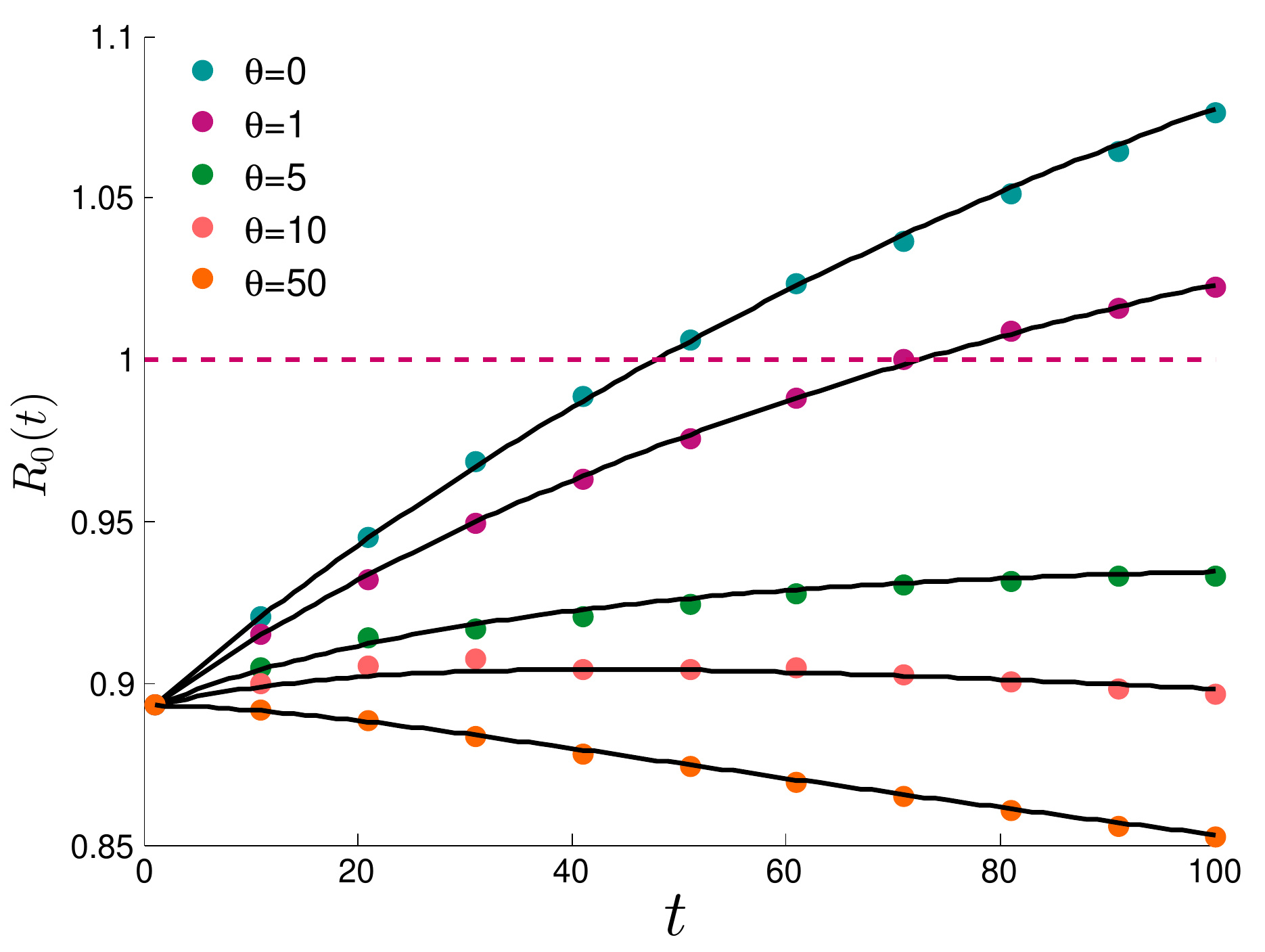}
\caption[Figure ]%
{ (Color online) Temporal evolution of $\mathcal{R}_0$ in the SIR model for the social network of dolphins. 
The initial connectivity of incoming nodes   is $\beta=2$. The transmission rate is $\eta=0.04$ and the recovery rate is $\mu=0.22$.
The solid markers are averaged over 100 Monte Carlo trials. The solid line represents the theoretical prediction~\eqref{R0_SIR_t}. 
The figure indicates that the higher the value of $\theta$ (initial attractiveness), the more will be the future  resilience of the network against endemic outbreaks. This means that less values of $\theta$, which is tantamount with more weight to be given by incoming nodes to the degrees of existing nodes, the more susceptible the network will be against endemic outbreaks. 
}
\label{Z_dolphin_R0_SIR_recov_22sadom_trans_4sadom_beta_2_theta_avaz}
\end{figure}

For the third case study, we consider the social network of dolphins.  We consider a growth process with $\theta=5$, and   investigate $\mathcal{R}_0(t)$ for different values of  $\beta$. 
 The results are depicted in Figure~\ref{Z_dolphin_R0_beta_avaz_th_5}. As the figure illustrates, increasing the value of $\beta$ increases the rate of increase  in $\mathcal{R}_0(t)$. This means that the more gregarious the new incoming individuals are, the more they will drive the network towards the outbreak threshold (i.e., $\mathcal{R}_0=1$). Furthermore, Figure~\ref{Z_dolphin_R0_SIR_recov_22sadom_trans_4sadom_beta_2_theta_avaz} illustrates the results for  fixed connectivity of incoming nodes, $\beta=2$, for different values of $\theta$. It can be observed that higher values of $\theta$ are better for the system in the sense that they decrease the basic reproduction number in time. This means that the more \emph{preferential} the growth mechanism is, the more susceptible the network will be against future endemic outbreaks. This is intuitively expected, because lower values of $\theta$ are closer to the conventional preferential attachment, for which it is known that the epidemic threshold vanishes  in the long-time limit.

%
%
%
%
%
%
%

\section{Summary and Discussion}
The first contribution of this paper is providing an analytical expression for the degree distribution of a growing network as a function of time, where the initial network on top of which the growth takes place can be arbitrary. The growth mechanism considered in this paper was shifted-linear growth, where nodes are endowed with initial attractiveness. We corroborate the theoretical findings with Monte Carlo simulations. 

The time-dependent degree-distribution is then used to analyze the epidemic threshold and the basic reproduction number of growing networks as a function of time. The results are shown to be in good agreement with simulations. We observe that as the new incoming nodes are added to the network, the epidemic threshold and the basic reproduction number change. Hence, new nodes can increase or decrease $\mathcal{R}_0$, which depends on the growth parameters (i.e., initial attractiveness and the initial connectivity of newcomers). This means that if a network whose $\mathcal{R}_0$ is below unity (hence resilient against endemic outbreaks) grows by the addition of new nodes to the system, $\mathcal{R}_0$ can be elevated above unity, changing the state of the epidemic. The reverse can also occur. The results indicate that   opposing effects from two sources compete on altering the value  of  $\mathcal{R}_0$  in time. The first source is $\theta$, the  initial attractiveness incorporated into the growth mechanism. As $\theta$ increases, the growth mechanism is more agnostic to the degrees of destination nodes, hence it is less preferential. Increasing $\theta$ increases the resilience of the network against endemic outbreaks. The second source is $\beta$, which is the number of initial connections that each newcomer establishes. Increasing $\beta$ renders the network more susceptible to endemic outbreaks. 

These results shed light on the impact of  population growth and migration  processes on the epidemic character of a networked society, and takes a step towards the prediction of such impact.

A plausible extension of the problem analyzed in this  paper  is to generalize it to networks with nonnegligible degree-degree correlations, for which the epidemic threshold obtained via the annealed mean-field approximation depends on the spectral properties of the branching matrix~\cite{RevModPhys.87.925} (also called the connectivity matrix~\cite{boguna2003epidemic,youssef2011individual}). To that end, the rate equation for the evolution of the degree correlations must be solved, so that the result can be used to construct the time-dependent branching matrix. Moreover, for regions with high levels of migration,  there is evidence that migration processes are partly driven by the network of connections between prospective migrants and those who have already migrated~\cite{bauer2000migration,bauer2002herd,haug2008migration,mckenzie2010self}, which leads to a self-perpertuating  flow of migrants~\cite{massey1987social}. Using the existing data from the migration networks can help develop models for how the network grows, and consequently, predict the concomitant changes in the epidemic properties.

Moreover, a more realistic depiction of the disease spread can ensue if one moves beyond the adiabatic approximation considered in the present paper, and study the interplay between disease dynamics and network growth in the case where time scales of the two dynamics are comparable.

\section{Acknowledgements}
The authors thank Naghmeh Momeni and Ardalan Khazraei for their fruitful suggestions and constructive comments towards the improvement of this manuscript.

%
%
%
%

\appendix
\section{Solving the PDE in~\eqref{psi_dot_1}}\label{app:pde}
The PDE we need to solve is:
\all{
\resizebox{.95 \linewidth}{!}{$
\DD \rond{\psi(z,t)}{t} - \fracc{\beta }{\zeta + \nu t} (z-1) \rond{\psi(z,t)}{z} = \fracc{\beta \theta}{\zeta + \nu t} (z^{-1}-1) \psi(z,t) + z^{-\beta}
$}
.}{psi_dot_1_app}
We employ the method of characteristic curves to solve this equation (see for example~\cite{zwillinger1998handbook}, for background on this method). We need to first solve the following system of equations:
\all{
\fracc{dt}{1}= \fracc{dz}{- \fracc{\beta }{\zeta + \nu t} (z-1)}
= \fracc{d \psi}{\fracc{\beta \theta}{\zeta + \nu t} (z^{-1}-1) \psi(z,t) + z^{-\beta}}
.}{sys}

From the first equation we get
\all{
\fracc{dt}{1}= \fracc{dz}{- \fracc{\beta }{\zeta + \nu t} (z-1)}
\Longrightarrow 
(z-1)^{\frac{\nu}{\beta }} (\zeta + \nu t) = C
.}{sys1}

The second equation is
\all{
\fracc{dz}{- \fracc{\beta }{\zeta + \nu t} (z-1)}
= \fracc{d \psi}{\fracc{\beta \theta}{\zeta + \nu t} (z^{-1}-1) \psi(z,t) + z^{-\beta}}
.}{sys2}

This can be rearranged and rewritten as follows
\all{
\fracc{d\psi}{dz}- \frac{\theta}{z} \psi = \fracc{- z^{-\beta} (\zeta+\nu t)}{\beta (z-1)}
.}{sys2_3}

Using~\eqref{sys1}, this transforms into

\all{
\fracc{d\psi}{dz}- \frac{\theta}{z} \psi = \fracc{- C z^{-\beta} }{\beta }(z-1)^{\frac{-\nu}{\beta }-1}
.}{sys2_3}

This is an ordinary first-order linear differential equation, with integrating factor $z^{-\theta}$. The solution is given by
\all{
\psi=z^{\theta} \left[ \fracc{- C }{\beta }\DD \int^z z'^{-\beta-\theta} (z'-1)^{\frac{-\nu}{\beta }-1} dz' + \Phi(C) \right] 
,}{psi_sol_1}

where $\Phi(C)$, according to the method of characteristics, is an arbitrary function of $C$ that is uniquely specified for given initial conditions. We expand the integrand before performing the integration. We have
\all{
(z-1)^{\frac{-\nu}{\beta }-1} =\DD z^{\frac{-\nu}{\beta }-1} \DD \sum_{m=0}^{\infty}  \binom{\frac{ \nu}{\beta }+m }{m} z^{-m}
.}{tylor_1}
Plugging this into~\eqref{psi_sol_1}, we get
\all{
\psi(z,t)&
\resizebox{.85\linewidth}{!}{$
=z^{\theta} \left[ \fracc{- C }{\beta } \DD \sum_{m=0}^{\infty} \binom{\frac{ \nu}{\beta }+m }{m}  \int^z z'^{\frac{-\nu}{\beta }-1-m-\beta-\theta} dz'
+\Phi(C) \right]
$}
\nonumber \\ 
&
=z^{\theta} \left[ \fracc{+ C }{\beta } \DD \sum_{m=0}^{\infty} \binom{\frac{ \nu}{\beta }+m }{m} \fracc{ z^{\frac{-\nu}{\beta }-m-\beta-\theta}}{\frac{\nu}{\beta } +m+\beta+\theta}
+\Phi(C) \right]
\nonumber \\ 
&
= \fracc{ C }{\beta } \DD \sum_{m=0}^{\infty}   \binom{\frac{ \nu}{\beta }+m }{m} \fracc{ z^{\frac{-\nu}{\beta }-m-\beta}}{\frac{\nu}{\beta } +m+\beta+\theta}
+\Phi(C)z^{\theta}
.}{psi_7}
Now we use~\eqref{sys1} to plug in the explicit expression for $C$ into~\eqref{psi_7}:
Plugging this into~\eqref{psi_sol_1}, we get
\all{
\psi(z,t) 
= &
\resizebox{.85\linewidth}{!}{$
\fracc{  (z-1)^{\frac{\nu}{\beta }} (\zeta + \nu t)  }{\beta } \DD \sum_{m=0}^{\infty}   \binom{\frac{ \nu}{\beta }+m }{m}  \fracc{ z^{\frac{-\nu}{\beta }-m-\beta}}{\frac{\nu}{\beta } +m+\beta+\theta}
$}\nonumber \\ &
+\Phi \left[ (z-1)^{\frac{\nu}{\beta }} (\zeta + \nu t) \right] z^{\theta}
.
}{psi_8}

Let us define
\all{
F(z ) \stackrel{\text{def}}{=} \fracc{ (z-1)^{\frac{\nu}{\beta }} }{\beta } \DD \sum_{m=0}^{\infty}   \binom{\frac{ \nu}{\beta }+m }{m}\fracc{ z^{\frac{-\nu}{\beta }-m-\beta}}{\frac{\nu}{\beta } +m+\beta+\theta}
}{F_def}

Then~\eqref{psi_8} can be rewritten as follows: 
\all{
\psi(z,t) 
= (\zeta+\nu t) F(z ) 
+ z^{\theta}\Phi \left[ (z-1)^{\frac{\nu}{\beta }} (\zeta + \nu t) \right]
.}{psi_9}

We need to uniquely determine $\Phi(\cdot)$. At time $t=0$, Equation~\eqref{psi_9} becomes
\all{
& \psi(z,0) 
=\zeta F(z ) 
+ z^{\theta} \Phi \left[ (z-1)^{\frac{\nu}{\beta }} \zeta \right]\Longrightarrow \nonumber \\ &
\Phi \left[ (z-1)^{\frac{\nu}{\beta }} \zeta \right] =z^{-\theta} \bigg[\psi(z,0)- \zeta F(z )\bigg]
\Longrightarrow \nonumber \\&
\Phi(X)=\left[ \left(\frac{X}{\zeta}\right)^{\frac{\beta }{\nu}}+1\right]^{-\theta}
\nonumber \\ & \times
\bigg[\psi \left( \left(\frac{X}{\zeta}\right)^{\frac{\beta }{\nu}}+1 , 0\right) - \zeta F \left( \left(\frac{X}{\zeta}\right)^{\frac{\beta }{\nu}}+1 \right) \bigg]
.
}{psi_10}
Also, let us define
\all{
c \stackrel{\text{def}}{=} 1 - \left( \fracc {\zeta}{\zeta+\nu t}\right)^{\frac{\beta }{\nu}}
.}{c_def}
Then it follows that
\all{
(z-1) \left(\frac{\zeta+\nu t}{\zeta}\right)^{\frac{\beta }{\nu}}+1=\fracc{z-c}{1-c}
.}{jaleb}
Using~\eqref{jaleb},~\eqref{c_def}, we can simplify~\eqref{psi_10} into the following:
\all{
&
z^{\theta} \Phi \left[ (z-1)^{\frac{\nu}{\beta }} (\zeta + \nu t) \right] 
= 
\nonumber \\ &
z^{\theta}
\left(\fracc{z-c}{1-c}\right)^{-\theta}
\bigg[\psi \left( \fracc{z-c}{1-c} , 0\right) - \zeta F \left(\fracc{z-c}{1-c} \right) \bigg]
}{phi_final}
Substituting the last term on the right hand side of~\eqref{psi_9} with the expression in~\eqref{phi_final}, we arrive at
\all{
& \psi(z,t) 
= (\zeta+\nu t) F(z) \nonumber \\ &
+ z^{\theta}
\left(\fracc{z-c}{1-c}\right)^{-\theta}
\bigg[\psi \left( \fracc{z-c}{1-c} , 0\right) -\zeta F \left(\fracc{z-c}{1-c} \right) \bigg]
.}{psi_9}

\section{Taking the Inverse Transform of~\eqref{psi_sol}} \label{app:inv}
Now we need to take the inverse transform of this expression. We do this term by term. First, we take the inverse transform of $F(z)$. We have

\all{
&F(z ) = \fracc{ (z-1)^{\frac{\nu}{\beta }} }{\beta } \DD \sum_{m=0}^{\infty}  \binom{\frac{ \nu}{\beta }+m }{m} \fracc{ z^{\frac{-\nu}{\beta }-m-\beta}}{\frac{\nu}{\beta } +m+\beta+\theta}
\nonumber \\ & 
= \fracc{ z^{\frac{\nu}{\beta }} (1-z^{-1})^{\frac{\nu}{\beta }} }{\beta } 
\DD \sum_{m=0}^{\infty}   \binom{\frac{ \nu}{\beta }+m }{m} \fracc{ z^{\frac{-\nu}{\beta }-m-\beta}}{\frac{\nu}{\beta } +m+\beta+\theta}
\nonumber \\ &
=
\fracc{ (1-z^{-1})^{\frac{\nu}{\beta }} }{\beta } 
\DD \sum_{m=0}^{\infty}  \binom{\frac{ \nu}{\beta }+m }{m} \fracc{ z^{-m-\beta}}{\frac{\nu}{\beta } +m+\beta+\theta}
\nonumber \\ &
=
\fracc{1 }{\beta } 
\DD \sum_{m,r} (-1)^{ r} \fracc{z^{-m-\beta }}{\frac{\nu}{\beta } +m-r+\beta+\theta}
\binom{\frac{ \nu}{\beta }+m-r }{m-r} \binom{\frac{ \nu}{\beta }}{r}.
}{F_inv_0}

The inverse Z-transform of $z^{-a}$ for some integer $a$ is $\delta[k-a]$. So we take the inverse transform of $F(z)$ term by term:

\all{
&
F(z ) \xrightarrow{\mathcal{Z}^{-1} } \nonumber\\ 
&
\fracc{1 }{\beta } 
\DD \sum_{m,r} (-1)^{ r} \fracc{\delta[k-m-\beta]}{\frac{\nu}{\beta } +m-r+\beta+\theta}
\binom{\frac{ \nu}{\beta }+m -r}{m-r} \binom{\frac{ \nu}{\beta }}{r}
\nonumber \\ 
=
&
\fracc{1 }{\beta } 
\DD \sum_{r} \fracc{ (-1)^{r}}{\frac{\nu}{\beta } +k+\theta-r}
\binom{\frac{ \nu}{\beta }+k-\beta -r}{k-\beta-r} \binom{\frac{ \nu}{\beta }}{r}
}{F_inv_1}

Now we utilize the following identity: 
\all{ 
&\DD \sum_{r} \fracc{ (-1)^{r}}{\frac{\nu}{\beta } +k+\theta-r}
\binom{\frac{ \nu}{\beta }+k-\beta -r}{k-\beta-r} \binom{\frac{ \nu}{\beta }}{r}
\nonumber \\ &
= 
\fracc{\Gamma(k+\theta)\Gamma(\beta+\frac{\nu}{\beta}+\theta)}{\Gamma(\beta+\theta)\Gamma(k+1+\frac{\nu}{\beta}+\theta)} 
}{iden_1}

The proof of this identity is  given in Appendix~\ref{app:iden_proof}. Using this result to perform the summation in~\eqref{F_inv_1}, we obtain
\all{
F(z ) \xrightarrow{\mathcal{Z}^{-1} }
\fracc{1 }{\beta } 
\fracc{\Gamma(k+\theta)\Gamma(\beta+\frac{\nu}{\beta}+\theta)}{\Gamma(\beta+\theta)\Gamma(k+1+\frac{\nu}{\beta}+\theta)} 
.}{F_inv}

This yields the inverse transform of the first term on the right hand side of~\eqref{psi_9}. For the second and third terms, we first ask: if the inverse transform of some function $F(z)$ is known, and is given by, say, $f_k$, then what is the inverse transform of $z^{\theta} \left(\frac{z-c}{1-c}\right)^{-\theta} F\left(\frac{z-c}{1-c}\right)$? We have: 
\all{
&z^{\theta} \left(\frac{z-c}{1-c}\right)^{-\theta} F\left(\frac{z-c}{1-c}\right) 
\nonumber \\ &
= 
(1-c)^{\theta} (z-c)^{-\theta} z^{\theta}\DD \sum_{r} f_r \left(\frac{z-c}{1-c}\right)^{-r}
\nonumber \\ &
\resizebox{\linewidth}{!}{$
=
(1-c)^{\theta} z^{\theta}\DD \sum_{r} \fracc{ f_r (1-c)^r }{ ( z-c )^{r+\theta}}
=(1-c)^{\theta} \DD \sum_{r} \fracc{ f_r z^{-r}(1-c)^r }{ ( 1-c z^{-1} )^{r+\theta}}
$}
\nonumber \\ &
=(1-c)^{\theta} \DD \sum_{r} \sum_j f_r z^{-r}(1-c)^r c^j \binom{r+\theta+j-1}{j} z^{-j}
\nonumber \\ &
=(1-c)^{\theta} \DD \sum_{k} \Bigg[ \sum_r f_r (1-c)^r c^{k-r} \binom{k+\theta-1 }{k-r } \Bigg]z^{-k}
.}{steps}
So the inverse transform of $z^{\theta} \left(\frac{z-c}{1-c}\right)^{-\theta} F\left(\frac{z-c}{1-c}\right)$ is given by
\all{
&z^{\theta} \left(\frac{z-c}{1-c}\right)^{-\theta} F\left(\frac{z-c}{1-c}\right)
\Longrightarrow 
\nonumber \\ &
c^k (1-c)^{\theta} \sum_{r } f_r \left(\frac{1-c}{c}\right)^r \binom{k+\theta-1}{r+\theta-1}
.}{inv_big}
Using this result, we can take the inverse transform of the other two terms on the right hand side of~\eqref{psi_9}. We obtain

\begin{widetext}
\all{
M_k(t)&
= 
(1-c)^{\theta} c^k \DD \sum_{r=1} ^{k} N_r(0) \left( \fracc{1-c}{c} \right)^r \CC{k+\theta-1}{r+\theta-1}
+
\DD 
\fracc{ \bigg[\zeta+ (2\beta +\theta) t\bigg] }{ \beta } 
\fracc{\Gamma(k+\theta) }{ \Gamma(\beta+\theta)} \fracc{\Gamma \left( \beta+ 2+\frac{\theta}{\beta} +\theta \right) }
{\Gamma \left( k+3+ \frac{\theta}{\beta} + \theta \right) } u(k-\beta)
\nonumber \\
&
- \fracc{ \zeta (1-c)^{\theta} c^k }{ \beta } 
\Gamma \left( \beta+ 2+\frac{\theta}{\beta} +\theta \right)
\fracc{\Gamma(k+\theta) }{ \Gamma(\beta+\theta)}
\DD \sum_{r=\beta} ^{k}
\displaystyle \frac{ \left( \fracc{1-c}{c} \right)^r }{(k-r)! \Gamma \left( r+3+ \frac{\theta}{\beta} + \theta \right)}.
}{Mk_FIN}
\end{widetext}


\section{The time-continuous approximation}
Now we focus on the error of the approximation made by assuming time to be continuous, as done in~\eqref{Nk_dot}. 
Let us repeat the equation for convenience of reference: 
\all{
\resizebox{.95\linewidth}{!}{$
\DD \rond{M_k(t)}{t} = \beta \left[\fracc{(k-1+\theta)M_{k-1}(t)-(k+\theta)M_k(t)}{\zeta + \nu t} \right]
+ \delta_{k,\beta}.
$}
}{Mk_dot} 

Let us use the Taylor expansion of $M_k(t)$, around timestep $n$ (for which we have $t=n$), in the interval ${t \in [n,n+1]}$. The Taylor theorem states that for some ${  \xi_n \in [n,n+1]}$, we have: 
\all{
M_k(t)=
&M_k(n)+ \left. \DD \rond{M_k(t)}{t}\right|_{t=n} (t-n) 
\nonumber \\ &
+ \left. \DD \rondd{M_k(t)}{t}\right|_{t=\xi_n} \fracc{(t-n)^2}{2}
.}{taylor_1}

Rewriting this equation at $t=n+1$, and employing~\eqref{Mk_dot} to express $\left. \rond{M_k(t)}{t}\right|_{t=n} $, we get

\all{
&M_k(n+1)-
M_k(n) = \delta_{k,\beta}
\nonumber \\ &
+ 
\beta \left[\fracc{(k-1+\theta)M_{k-1}(n)-(k+\theta)M_k(n)}{\zeta + \nu n} \right]
\nonumber \\ &
+ \left. \fracc{1}{2}\DD \rondd{M_k(t)}{t} \right|_{t=\xi_n} 
.}{taylor_1}

Combining this with~\eqref{Nk_dot_0}, we arrive at the following recurrence for the error ${E_k(n)=M_k(n)-N_k(n)}$ with the boundary condition ${E_k(0)=0~ \forall k}$: 

\all{
&E_k(n+1)-
E_k(n) = 
\nonumber \\ &
\beta \left[\fracc{(k-1+\theta)E_{k-1}(n)-(k+\theta)E_k(n)}{\zeta + \nu n} \right]
\nonumber \\ &
+ \left. \fracc{1}{2}\DD \rondd{M_k(t)}{t} \right|_{t=\xi_n} 
.}{E1}

\begin{widetext}
If we take the time-derivative of both sides of~\eqref{Mk_dot}, and then employ~\eqref{Mk_dot} itself to express the first time derivatives appearing on the right hand side, and after straightforward algebraic steps, we arrive at the following expression for the second time derivative of $M_k(t)$:
\all{
&\DD \rondd{M_k(t)}{t}= \fracc{\beta}{\zeta+\nu t} \bigg[ (k+\theta-1) \delta_{k,\beta+1} - (k+\theta) \delta_{k,\beta} \bigg] 
\nonumber \\ 
&
\resizebox{\linewidth}{!}{$ +
\fracc{\beta^2}{(\zeta + \nu t)^2} \left[ (k+\theta-1)(k+\theta-2) M_{k-2}(t) 
- 2 (k+\theta-1) \left( k+\theta -\fracc{1}{2}+ \fracc{\nu}{2\beta}\right) M_{k-1}(t)
+ (k+\theta) \left(k+\theta+\fracc{\nu}{\beta} \right) M_{k}(t)
\right]
$}
}{Mkddot}
\end{widetext}

Now let us find an upper bound for $ \fracc{1}{2}\DD \rondd{M_k(t)}{t} $. Since it has $M_k(t)$ terms in it, we need bounds for terms in $M_k(t)$ as they appear on the right hand side of~\eqref{Mk_FIN}. For the first term, note that we have
\al{ 
& \sum_{r=1}^k (1-c)^{r+\theta} c^{k-r} \CC{k+\theta-1}{r+\theta-1}
\nonumber \\ &
\leq
\sum_{r=1}^k (1-c)^{r+\theta} c^{k-r} \CC{k+\ceil{\theta}-1}{r+\ceil{\theta}-1}
\nonumber \\ &
=
(1-c)^{\theta-\ceil{\theta}+1} \sum_{r=1}^k (1-c)^{r+\ceil{\theta}-1} c^{k -r} \CC{k+\ceil{\theta}-1}{r+\ceil{\theta}-1}
\nonumber \\ &
=
(1-c)^{\theta-\ceil{\theta}+1} \sum_{r'=\ceil{\theta}}^{k+\ceil{\theta}-1} (1-c)^{r'} c^{k+\ceil{\theta}-1-r'} \CC{k+\ceil{\theta}-1}{r'}
\nonumber \\ &
\leq
(1-c)^{\theta-\ceil{\theta}+1} \sum_{r'=0}^{k+\ceil{\theta}-1} (1-c)^{r'} c^{k+\ceil{\theta}-1-r'} \CC{k+\ceil{\theta}}{r'}
\nonumber \\ &
=
(1-c)^{\theta-\ceil{\theta}+1} \times 1=(1-c)^{\theta-\floor{\theta}} \leq 1
}

So for the first term on the right hand side of~\eqref{Mk_FIN}, we have
\all{
&
(1-c)^{\theta} c^k \DD \sum_{r=1} ^{k} N_r(0) \left( \fracc{1-c}{c} \right)^r \CC{k+\theta-1}{r+\theta-1}
\nonumber \\ &
\leq
\sum_{r=1}^k (1-c)^{r+\theta} c^{k-r} \CC{k+\theta-1}{r+\theta-1}
\times 
\sum_{r=1} ^{k} N_r(0) 
\nonumber \\ &
\leq 
\sum_{r=1} ^{k} N_r(0) \leq N(0)
.}{rhs1}
For the second term on the right hand side of~\eqref{Mk_FIN}, we have 
\all{
&
\DD 
\fracc{ (\zeta+ \nu t) }{ \beta } 
\fracc{\Gamma(k+\theta) }{ \Gamma(\beta+\theta)} \fracc{\Gamma \left( \beta+ 2+\frac{\theta}{\beta} +\theta \right) }
{\Gamma \left( k+3+ \frac{\theta}{\beta} + \theta \right) } u(k-\beta)
\nonumber \\ &
= \fracc{ (\zeta+ \nu t) }{ \left( k+2+ \frac{\theta}{\beta} + \theta \right) \beta } 
\fracc{\CC{\beta+\theta+1+\frac{\theta}{\beta}}{\beta+\theta}}
{\CC{k+\theta+1+\frac{\theta}{\beta}}{k+\theta}} u(k-\beta)
\nonumber \\ &
\leq
\fracc{ (\zeta+ \nu t) }{ \left( k+2+ \frac{\theta}{\beta} + \theta \right) \beta } u(k-\beta)
.
}{rhs2}

\begin{widetext}
For the third term on the right hand side of~\eqref{Mk_FIN}, we have 
\all{
&\DD \sum_{r=\beta} ^{k}
\displaystyle \frac{(1-c)^{\theta+r}c^{k-r} \Gamma \left( \beta+ 2+\frac{\theta}{\beta} +\theta \right)\Gamma(k+\theta) }{(k-r)! \Gamma \left( r+3+ \frac{\theta}{\beta} + \theta \right) \Gamma(\beta+\theta)}
=
\DD \sum_{r=\beta} ^{k}
(1-c)^{r+\theta} c^{k-r}
\CC{k+\theta-1}{r+\theta-1}
\fracc{ \Gamma (r+\theta) \Gamma \left( \beta+ 2+\frac{\theta}{\beta} +\theta \right)}
{\Gamma \left( r+3+ \frac{\theta}{\beta} + \theta \right) \Gamma(\beta+\theta)}
\nonumber \\ &
=
\DD \sum_{r=\beta} ^{k}
(1-c)^{r+\theta} c^{k-r}
\CC{k+\theta-1}{r+\theta-1}
\fracc{ \CC{\beta+\theta+1+\frac{\theta}{\beta}}{\beta+\theta-1}}{\CC{r+\theta+1+\frac{\theta}{\beta}}{r+\theta-1}}
\frac{1}{r+\theta+2+\frac{\theta}{\beta}}
\leq
\DD \sum_{r=\beta} ^{k}
(1-c)^{r+\theta} c^{k-r}
\CC{k+\theta-1}{r+\theta-1}
\fracc{ \CC{\beta+\theta+1+\frac{\theta}{\beta}}{\beta+\theta-1}}{\CC{r+\theta+1+\frac{\theta}{\beta}}{r+\theta-1}}
\nonumber \\ &
\leq
\DD \sum_{r=\beta} ^{k}
(1-c)^{r+\theta} c^{k-r}
\CC{k+\theta-1}{r+\theta-1} \leq 1
}{rhs3}
\end{widetext}

Combining~\eqref{rhs1},~\eqref{rhs2}, and~\eqref{rhs3}, we find the following upper bound for $M_k(t)$: 
\all{
|M_k(t)| \leq N(0) 
+ \fracc{ (\zeta+ \nu t) }{ \left( k+2+ \frac{\theta}{\beta} + \theta \right) \beta } u(k-\beta)
+ \fracc{\zeta}{\beta}
.}{b1}

\begin{widetext}
Let us denote $\theta+2+\ceil{\frac{\theta}{\beta}}$ by $\phi$. This means that for the second derivative of $M_k$, as given in~\eqref{Mkddot}, we have: 
\all{
\frac{1}{2} \left| \DD \rondd{M_k(t)}{t} \right|
&\leq 
\fracc{\beta(k+\phi)}{2(\zeta+\nu t)} \bigg[ \delta_{k,\beta+1} + \delta_{k,\beta} \bigg] 
+
\fracc{2 \beta^2 (k+\phi)^2}{(\zeta + \nu t)^2}
\left[N(0)+\fracc{\zeta+\nu t}{\beta (k+\phi-1)}u(k-\beta)+\frac{\zeta}{\beta}\right]
\nonumber \\ &
\leq
\fracc{\beta(k+\phi)}{\zeta+\nu t}  u(k-\beta)
+
\fracc{2 \beta^2 (k+\phi)^2}{(\zeta + \nu t)^2}
\left[N(0)+2\fracc{\zeta+\nu t}{\beta} u(k-\beta)\right]
\nonumber \\ &
\leq
\fracc{5\beta(k+\phi)^2}{\zeta+\nu t} u(k-\beta)
+\fracc{2 \beta^2 (k+\phi)^2}{(\zeta + \nu t)^2}N(0)
}{M2bound}
Equipped with this upper bound, we can now find an upper bound on the error. Let $\mathcal{E}_1 (n)$ denote the absolute value of the maximum error $E_k(n)$ over all $k$ at timestep $n$, that is, $\max_{k,k <\beta} |E_k(n)|$. 
Using~\eqref{E1}, and denoting $k_{\textnormal{max}}(0)+\phi$ by $K$, we have 
\all{
\mathcal{E}_1(n+1)
&=
\max_{k,k <\beta}
\left| E_k(n) + 
\beta \left[\fracc{(k-1+\theta)E_{k-1}(n)-(k+\theta)E_k(n)}{\zeta + \nu n} \right]
+ \left. \fracc{1}{2}\DD \rondd{M_k(t)}{t} \right|_{t=\xi_n} \right|
\nonumber \\ &
\resizebox{.88\linewidth}{!}{$
\leq 
\max_{k,k <\beta}
\left|E_k(n)  \left[ 1-\frac{\beta(k+\theta)}{\zeta+\nu n} \right]
+ 
 E_{k-1}(n)  \left[ \frac{\beta(k+\theta)}{\zeta+\nu n} \right]
 \right|
+
\max_{k,k <\beta}
\fracc{5\beta(k+\phi)^2}{\zeta+\nu n} u(k-\beta) +
\max_{k,k <\beta}
\fracc{2 \beta^2 (k+\phi)^2}{(\zeta + \nu n)^2}N(0)
$}
\nonumber \\ &
\leq 
\mathcal{E}_1(n) 
+\fracc{2 \beta^2 (\beta+\phi)^2}{(\zeta + \nu n)^2}N(0)
.}{E1}

This recursion yields the following bound on the error at timestep $n$: 
\all{
\mathcal{E}_1(n) 
&
\leq  
2 \beta^2 (\beta+\phi)^2 N(0) \DD \sum_{i=0}^n \fracc{1}{(\zeta+ \nu n)^2}
\leq 
 2 \beta^2 (\beta+\phi)^2 N(0)
\DD \int_{-1}^{n-1} 
\fracc{dt}{(\zeta+ \nu t)^2}
\nonumber \\ &
\leq 
 2 \beta^2 (\beta+\phi)^2 N(0)
\DD \int_{-1}^{n-1} 
\fracc{dt}{(\zeta+ \nu t)^2}
\leq 
 2 \beta^2 (\beta+\phi)^2 N(0)
\DD \int_{-1}^{n} 
\fracc{dt}{(\zeta+ \nu t)^2}
\nonumber \\ &
\leq 
 \fracc{2 \beta^2 (\beta+\phi)^2 N(0)}{\zeta - \nu}
\fracc{1+  n }{ \zeta + \nu n}
.}{Eb1}

The error grows over time, but reaches a constant at long times. At long times, $M_k(t)$ becomes small (because the majority of the network become the newcomers, and the effects of the initial network vanish), and at the limit as $t \rightarrow \infty$, all nodes in the network have degree greater than or equal to $\beta$. At this limit, $M_k(t)$  correctly tends to zero, as given by~\eqref{Mk_FIN}. For short times, the maximum error is inversely proportional to $\zeta^2$. Since $\zeta$  comprises the number of links in the initial network, the error is small. In intermediary times, it is not readily clear how the maximum error compares with the size of the solution itself. Monte Carlo simulations of various different topologies suggest that the error is negligibly small. 

Similarly, for $k \geq \beta$, we define $\mathcal{E}_2(n)$ to be the maximum  error. We have: 
\all{
\mathcal{E}_2(n+1)
&=
\max_{k,k \geq \beta}
\left| E_k(n) + 
\beta \left[\fracc{(k-1+\theta)E_{k-1}(n)-(k+\theta)E_k(n)}{\zeta + \nu n} \right]
+ \left. \fracc{1}{2}\DD \rondd{M_k(t)}{t} \right|_{t=\xi_n} \right|
\nonumber \\ &
\resizebox{.88\linewidth}{!}{$
\leq 
\max_{k,k  \geq \beta}
\left|E_k(n)  \left[ 1-\frac{\beta(k+\theta)}{\zeta+\nu n} \right]
+ 
 E_{k-1}(n)  \left[ \frac{\beta(k+\theta)}{\zeta+\nu n} \right]
 \right|
+
\max_{k,k <\beta}
\fracc{5\beta(k+\phi)^2}{\zeta+\nu n} u(k-\beta) +
\max_{k,k <\beta}
\fracc{2 \beta^2 (k+\phi)^2}{(\zeta + \nu n)^2}N(0)
$}
\nonumber \\ &
\leq 
\mathcal{E}_2(n) 
+
\fracc{5\beta K^2}{\zeta+\nu n} +\fracc{2 \beta^2 K^2}{(\zeta + \nu n)^2}N(0)
.}{E2}

This recursion yields the following bound on the error at timestep $n$: 
\all{
\mathcal{E}_2(n) 
&
\leq \DD 5 \beta K^2 \sum_{i=0}^n \fracc{1}{\zeta+ \nu n} 
+
2 \beta^2 K^2 N(0) \DD \sum_{i=0}^n \fracc{1}{(\zeta+ \nu n)^2}
\leq 
5 \beta K^2 \DD \int_{-1}^{n-1} \fracc{dt}{\zeta+ \nu t} 
+2 \beta^2 K^2 N(0)
\DD \int_{-1}^{n-1} 
\fracc{dt}{(\zeta+ \nu t)^2}
\nonumber \\ &
\leq 
5 \beta K^2 \DD \int_{-1}^{n-1} \fracc{dt}{\zeta+ \nu t} 
+2 \beta^2 K^2 N(0)
\DD \int_{-1}^{n-1} 
\fracc{dt}{(\zeta+ \nu t)^2}
\leq 
5 \beta K^2 \DD \int_{-1}^{n} \fracc{dt}{\zeta+ \nu t} 
+2 \beta^2 K^2 N(0)
\DD \int_{-1}^{n} 
\fracc{dt}{(\zeta+ \nu t)^2}
\nonumber \\ &
\leq 
\fracc{ 5 \beta K^2}{\nu} \log \left( 1 + \fracc{\nu n}{\zeta}\right)
+\fracc{2 \beta^2 K^2 N(0)}{\zeta - \nu}
\fracc{1+  n }{ \zeta + \nu n}
.}{Eb2}

For long times, $M_k(t)$ grows linearly, as given by~\eqref{Mk_FIN}. In this limit, the second term in~\eqref{Eb2} reaches a constant, and is negligible as compared to the logarithmic term. The logarithmic term grows in time, but its ratio to the linear growth that $M_k(t)$ undergoes, reaches zero, which means that the relative error reaches zero at long times. For short times, the argument of the logarithm is close to unity, hence the logarithmic term vanishes, and the second term of~\eqref{Eb2} prevails. This becomes similar to the case of  $k<\beta$ which is discussed above. For intermediary times, eyeballing does not provide a straightforward understanding for how the error grows as compared to the size of the solution itself. Our evidence for the negligibly-small error is the Monte Carlo simulations conducted on various diverse topologies.

\section{Proof the the Identity Given in Equation~\eqref{iden_1}}  \label{app:iden_proof}

Let us repeat the identity (and rewrite the binomial coefficients in extended form) for easy reference:
\all{
\DD \sum_{r=0}^{k-\beta} 
 \fracc{  (-1)^r \Gamma \left(1+\nu + k-r-\beta \right) }{r! (k-r-\beta)! \Gamma( 1+\nu-r )(k+2-r +\theta+\frac{\theta}{\beta} )}
= \fracc{(k+\theta-1)!}{(\beta+\theta-1)!} \fracc{\Gamma \left( \beta+2+\frac{\theta}{\beta}+\theta  \right) }
{\Gamma \left( k+1+\nu + \theta \right) }
.
}{iden_1_app}

We will denote the left hand side of this equality by $h_k$. Define
\all{
\mathcal{H}(x) \deff \DD \sum_k h_k x^k
.}{H_def_app}

Also, the following identities can be immediately proved through elementary Taylor expansions:
\al{
\DD \sum_{j=0}^{\infty} \fracc{\Gamma(j+\alpha)   }{\Gamma(\alpha) j!} x^j   &= (1-x)^{-\alpha}
\label{iden_pair_1} \\
\DD \sum_{j=0}^{\infty} \fracc{\Gamma( \alpha+1)   }{\Gamma(\alpha+1-j) j!} x^j   &= (1+x)^{ \alpha}  .
\label{iden_pair_2}
}
We will begin from the first hand side of~\eqref{iden_1_app}. We have: 

\all{
\mathcal{H}(x) &= \DD \sum_k \sum_r \fracc{\Gamma \left(1+\nu + k-r-\beta \right) }{  (k-r-\beta)!}
 \fracc{  (-1)^r }{r!\Gamma( 1+\nu-r )}
\fracc{1}{k+2-r +\theta+\frac{\theta}{\beta}} x^k 
\nonumber \\
&= \DD \sum_r  \fracc{  (-1)^r }{r!\Gamma( 1+\nu-r )}
\DD \sum_k 
 \fracc{\Gamma \left(1+\nu + k-r-\beta \right) }{  (k-r-\beta)!}
 \fracc{x^k}{k+2-r +\theta+\frac{\theta}{\beta}} 
 \nonumber \\
&= \DD \sum_r  \fracc{  (-1)^r }{r!\Gamma( 1+\nu-r )}
x^{-2+r-\theta-\frac{\theta}{\beta}}
\DD \sum_k 
 \fracc{\Gamma \left(1+\nu + k-r-\beta \right) }{  (k-r-\beta)!}
 \fracc{x^{k+2-r +\theta+\frac{\theta}{\beta}}}{k+2-r +\theta+\frac{\theta}{\beta}} 
  \nonumber \\
 &= \DD \sum_r  \fracc{  (-1)^r }{r!\Gamma( 1+\nu-r )}
x^{-2+r-\theta-\frac{\theta}{\beta}}
\DD \sum_k 
 \fracc{\Gamma \left(1+\nu + k-r-\beta \right) }{  (k-r-\beta)!}
\DD \int^x x^{k+1-r +\theta+\frac{\theta}{\beta}} dx
 \nonumber \\
 &= \DD \sum_r  \fracc{  (-1)^r \Gamma(1+\nu ) }{r!\Gamma( 1+\nu-r )}
x^{-2+r-\theta-\frac{\theta}{\beta}}
\DD \int^x\DD \sum_k 
 \fracc{\Gamma \left(1+\nu + k-r-\beta \right) }{\Gamma(1+\nu )  (k-r-\beta)!}
 x^{k+1-r +\theta+\frac{\theta}{\beta}} dx
  \nonumber \\
 &= \DD \sum_r  \fracc{  (-1)^r \Gamma(1+\nu ) }{r!\Gamma( 1+\nu-r )}
x^{-2+r-\theta-\frac{\theta}{\beta}}
\DD \int^x\DD
x^{1+\theta+\beta+\frac{\theta}{\beta}}
  \sum_k 
 \fracc{\Gamma \left(1+\nu + k-r-\beta \right) }{\Gamma(1+\nu )  (k-r-\beta)!}
 x^{k-r-\beta } dx
  \nonumber \\
 &\stackrel{\text{\eqref{iden_pair_1}}}{=}
  \DD \sum_r  \fracc{  (-1)^r \Gamma(1+\nu ) }{r!\Gamma( 1+\nu-r )}
x^{-2+r-\theta-\frac{\theta}{\beta}}
\DD \int^x\DD
\DD x^{1+\theta+\beta+\frac{\theta}{\beta}}
\DD (1-x)^{-3-\frac{\theta}{\beta}} dx
  \nonumber \\
 &=
\bigg[   \DD \sum_r  \fracc{  (-x)^r \Gamma(1+\nu ) }{r!\Gamma( 1+\nu-r )}
x^{-2 -\theta-\frac{\theta}{\beta}}\bigg] 
\DD \int^x\DD
\DD x^{1+\theta+\beta+\frac{\theta}{\beta}}
\DD (1-x)^{-3-\frac{\theta}{\beta}} dx
  \nonumber \\
 &\stackrel{\text{\eqref{iden_pair_2}}}{=}
(1-x)^{ 2+\frac{\theta}{\beta}}
x^{-2 -\theta-\frac{\theta}{\beta}}
\DD \int^x\DD
\DD x^{1+\theta+\beta+\frac{\theta}{\beta}}
\DD (1-x)^{-3-\frac{\theta}{\beta}} dx
.}{H_steps}

Let us define
\all{
\begin{cases}
f_1(x) &\deff  \DD (1-x)^{ 2+\frac{\theta}{\beta}}
x^{-2 -\theta-\frac{\theta}{\beta}}
 \\
f_2(x) &\deff \int^x\DD
\DD x^{1+\theta+\beta+\frac{\theta}{\beta}}
\DD (1-x)^{-3-\frac{\theta}{\beta}} dx
.
\end{cases}
}{f_1_2}

Then we can rewrite~\eqref{H_steps} in the following compact form: 
 $
\mathcal{H}(x)=f_1(x) f_2(x)
$. Taking the derivative of both sides, we get
\all{
\mathcal{H}'(x) &= f_1(x)f_2'(x) + f_1'(x)f_2(x)
= f_1(x)f_2'(x) + f_1'(x)\frac{\mathcal{H}(x)}{f_1(x)}
=f_1(x)f_2'(x) + \frac{f_1'(x)}{f_1(x)}\mathcal{H}(x) 
.}{Hp}

From the definition of $f_1(x)$ and $f_2(x)$, the following can be reached through elementary algebraic steps:
\all{
\begin{cases}
f_1(x)f_2'(x)& = (1-x)^{-1}  x^{\beta-1}\\
\frac{f_1'(x)}{f_1(x)}&=\fracc{ - \left(2+\theta+\frac{\theta}{\beta}  \right)  }{x(1-x)}+\fracc{\theta}{1-x}
 . 
\end{cases}
}{f1p_f1}

Using Taylor expansion, the following holds: 
\all{
\fracc{ - \left(2+\theta+\frac{\theta}{\beta}  \right)  }{x(1-x)}+\fracc{\theta}{1-x}
 =
\fracc{-\theta}{x} - \left(2 +\frac{\theta}{\beta}  \right)  \DD \sum_{k=-1}^{\infty}  x^k 
.}{taylor_f1}

Plugging  the expansion form of $\mathcal{H}(x)$ given in~\eqref{H_def_app} into~\eqref{Hp} and using~\eqref{taylor_f1} and~\eqref{f1p_f1}, we get
\all{
&\sum_k (k+1)h_{k+1} x^k   = (1-x)^{-1} x^{\beta-1}
+\left[
-\fracc{ \theta}{x} - \left(2 +\frac{\theta}{\beta}  \right)  \DD \sum_{k=-1}^{\infty}  x^k  \right] \sum_k h_k x^k 
 . 
}{hk_temp_1}
Equating the coefficients of $x^k$ on both sides, we get
\all{
(k+1)h_{k+1} &=u(k+1-\beta )-\theta h_{k+1}
- \left(2 +\frac{\theta}{\beta}  \right)  \sum_{j=0}^{k+1}  h_j
.}{coefs_1}

Let us write the same equation for $k$ rather than $k+1$:

\all{
(k )h_{k }  =u(k-\beta )-\theta h_{k }
- \left(2 +\frac{\theta}{\beta}  \right)  \sum_{j=0}^{k }  h_j   
.}{coefs_1_k}

Subtracting~\eqref{coefs_1_k} from~\eqref{coefs_1}, we get
 
\all{
(k+1)h_{k+1}-kh_k
&=\delta[k+1-\beta] -\theta h_{k+1} + \theta h_k 
 - \left(2 +\frac{\theta}{\beta}  \right) h_{k+1}
.}{subtract_hs}
This can be expressed equivalently as follows: 
\all{
 h_{k+1}=  &
\fracc{ (\theta+k)}{\left( k+3 + \theta + \frac{\theta}{\beta}  \right)}
  h_k 
  + \fracc{1}{\left( k+3 + \theta + \frac{\theta}{\beta} \right)} 
  \delta[k+1-\beta]
.}{hs_1}

Let us find the solution to  this recurrence relation.    For ${k=\beta-1}$ we have
\al{
h_{\beta} =  \fracc{1}{\left( \beta+2 + \theta + \frac{\theta}{\beta}  \right)} 
.}
For $k=\beta$ we have
$
h_{\beta+1} =  \frac{(\beta+\theta)}{\left( \beta+2 + \theta + \frac{\theta}{\beta} \right)\left( \beta+3 + \theta + \frac{\theta}{\beta}  \right)}
$. 
For $k=\beta+1$ we have $
h_{\beta+2} =  \frac{ (\beta+\theta) (\beta+\theta+1)}{\left( \beta+2 + \theta + \frac{\theta}{\beta} \right)\left( \beta+3 + \theta + \frac{\theta}{\beta} \right)\left( \beta+4 + \theta + \frac{\theta}{\beta}  \right)}
$. 
The pattern is apparent. For general $k$ we have
\all{
h_k
=
\fracc{\prod_{j=0}^{k -\beta-1} (\beta+\theta+j)}
{\prod_{j=0}^{k-\beta }  \left( \beta+2 + \theta + \frac{\theta}{\beta}  +j \right)}
.}{hk_semifinal}

The numerator equals $\frac{(k+\theta-1)!}{\beta+\theta-1)!}$. Using the properties of the Gamma function, namely the fact that ${\Gamma(x+1)=x \Gamma(x)}$, the denominator can be written as $\frac{\Gamma(k+2+\theta+\frac{\theta}{\beta})}{\Gamma(\beta+3+\theta+\frac{\theta}{\beta})}$. Pugging these two expressions into~\eqref{hk_semifinal}, we arrive at
\all{
h_k= 
  \fracc{(k+\theta-1)!}{(\beta+\theta-1)!} \fracc{\Gamma \left( \beta+2+\frac{\theta}{\beta}+\theta  \right) }
{\Gamma \left( k+1+\nu + \theta \right) }
.}{hk_final}

This is identical to the right hand side of~\eqref{iden_1_app}, hence,  the proof is concluded.

\end{widetext}

\section{More Simulation Results for the Temporal Evolution of the Degree Distribution}\label{app:Pk_t}

Here we provide more simulation results on the accuracy of the theoretical predictions for the degree distribution as a function of  time, as given in~\eqref{Pk_ultimate}.

\begin{figure}[h]
\centering
\includegraphics[width=.95 \columnwidth]{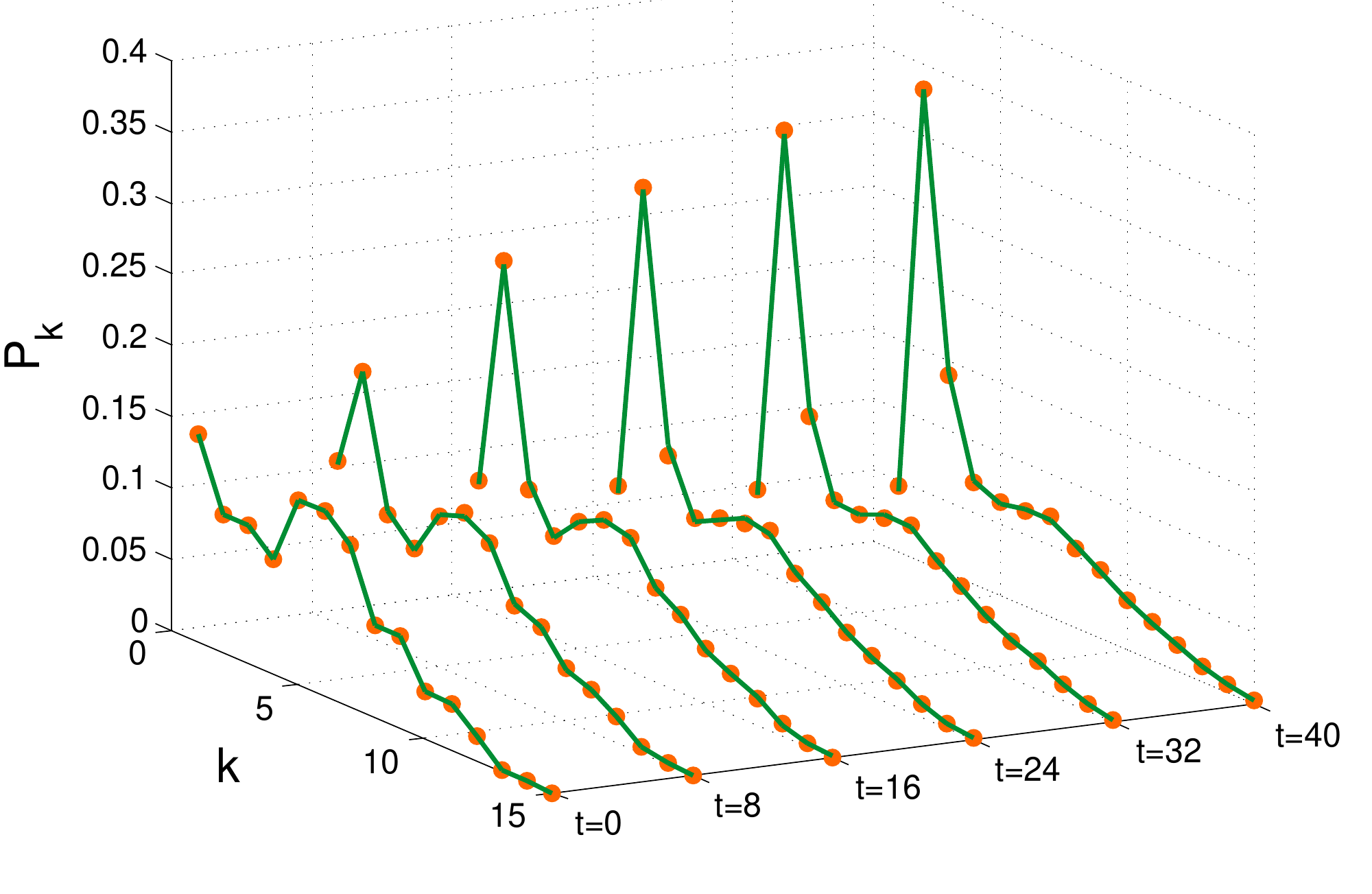}
\caption[Figure ]%
{ (Color online)  Theoretical prediction and the simulation results for the growth process, using the social network of dolphins as the substrate. The growth parameters are $\beta=2$ and $\theta=10$. 
}
\label{Z_dolphinPkt_beta_2_th_10}
\end{figure}

We first take the social network of dolphins as the substrate. The  parameters of the growth process are $\beta=2$ and $\theta=10$. The  substrate  comprises 62 nodes. Figure~\ref{Z_dolphinPkt_beta_2_th_10} depicts the results.

\begin{figure}[h]
\centering
\includegraphics[width=.95 \columnwidth]{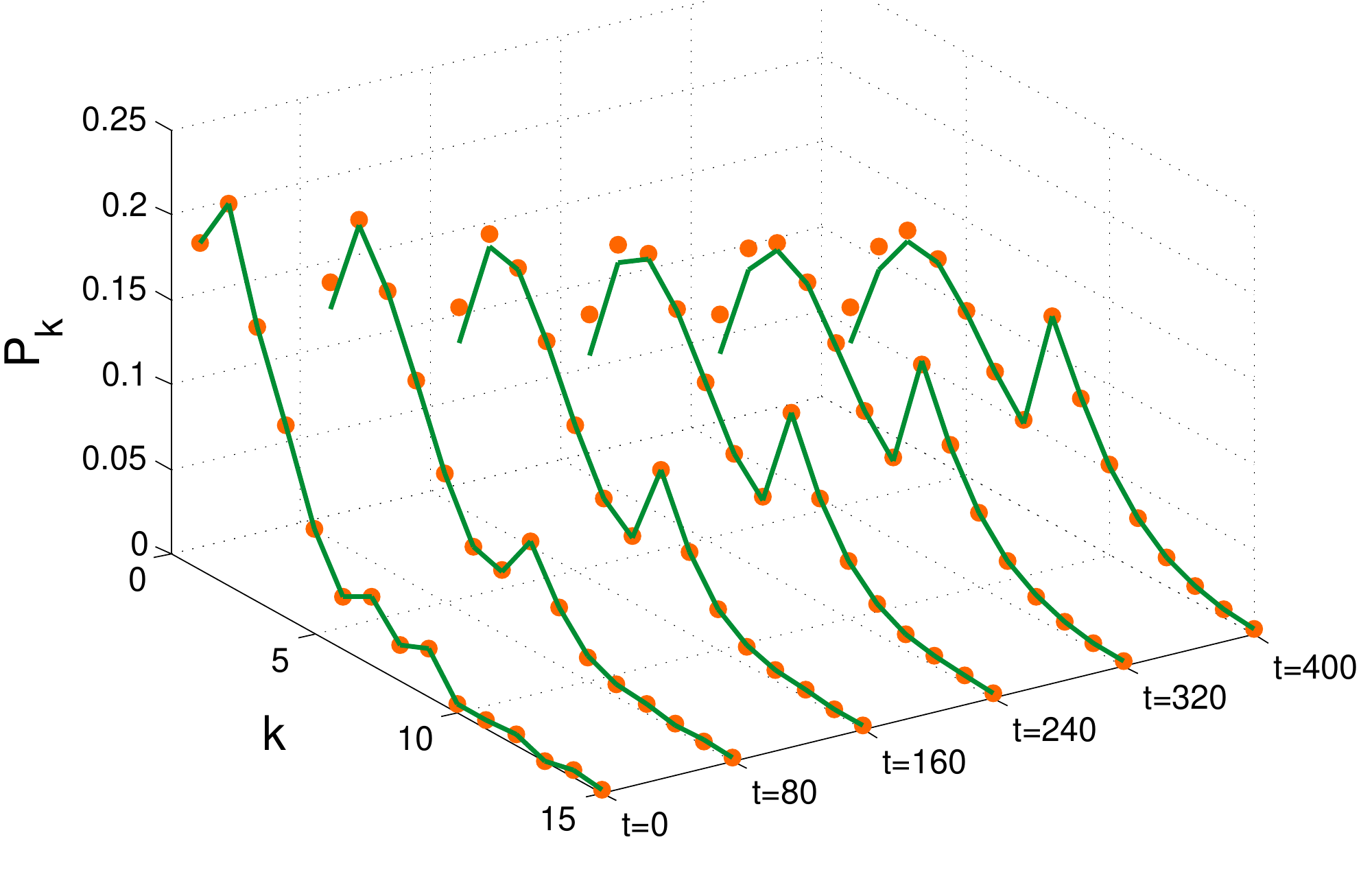}
\caption[Figure ]%
{ (Color online)  Theoretical prediction and the simulation results for the growth process, using the network of collaborations between scholars of network science as the substrate for the growth process. The growth parameters are $\beta=8$ and $\theta=9.5$.
}
\label{Z_netscience_Pkt_beta_8_th_10}
\end{figure}

The next network that we take as the substrate for the growth process is the network of collaborations among network science scholars. The growth parameters are $\theta=9.5$ and $\beta=8$. Figure~\ref{Z_netscience_Pkt_beta_8_th_10} illustrates the results.

\begin{figure}[h]
\centering
\includegraphics[width=.95 \columnwidth]{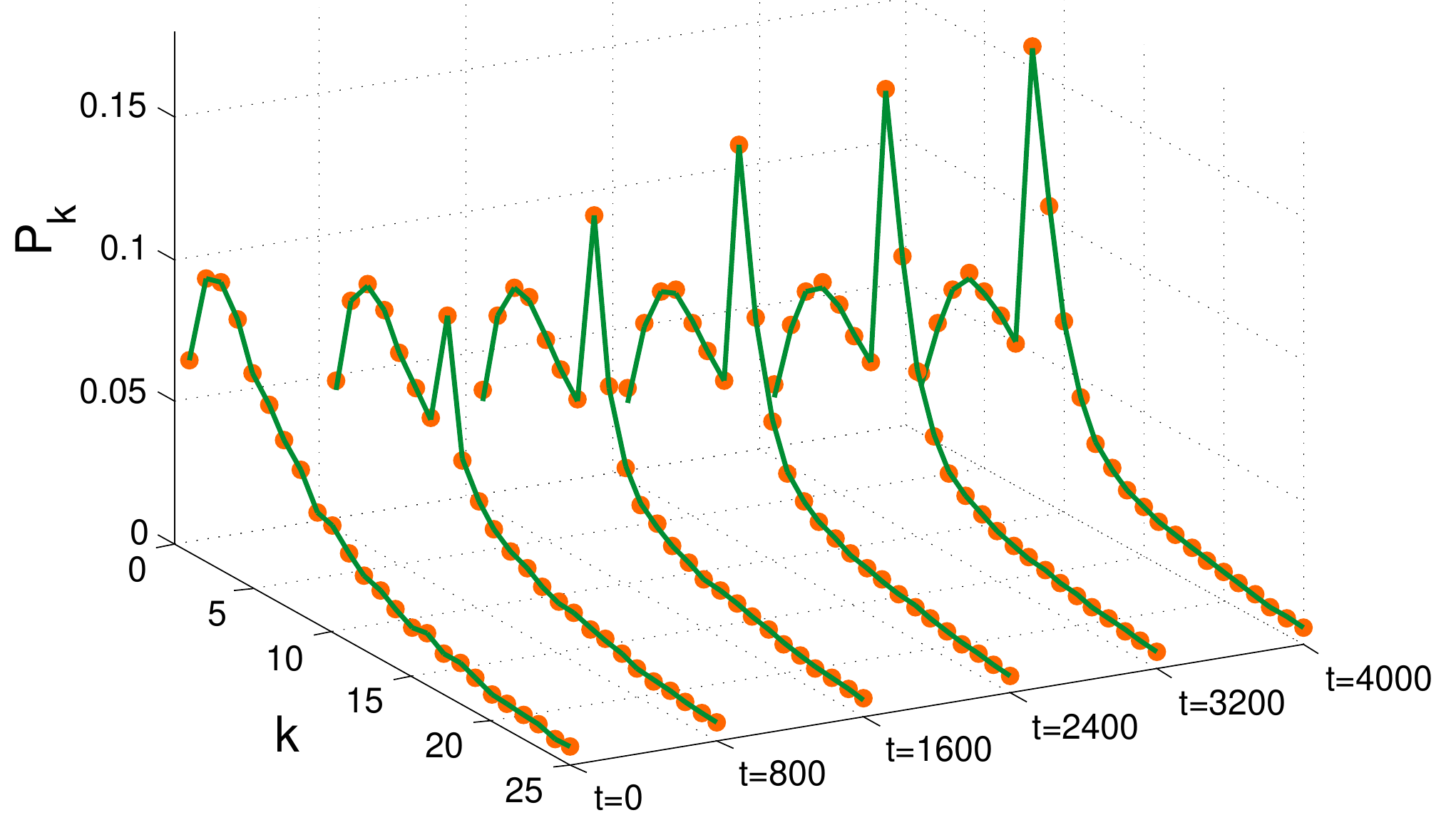}
\caption[Figure ]%
{ (Color online)  Theoretical prediction and the simulation results for the growth process, using the network of collaborations between  condensed matter physicists as the substrate for the growth process. The growth parameters are $\beta=8$ and $\theta=9.5$.
}
\label{Z_Pkt_condmat_beta_8_theta_10}
\end{figure}

Finally, we take the collaboration network of condensed matter physicists. The network has 13000 nodes. We consider $\beta=8$ and $\theta=9.5$ for the growth mechanism. The results are presented in Figure~\ref{Z_Pkt_condmat_beta_8_theta_10}.

\section{Simulation results for the epidemic spread over other topologies}\label{app:more_SIS}

\begin{figure}[h]
\centering
\includegraphics[width=.95 \columnwidth]{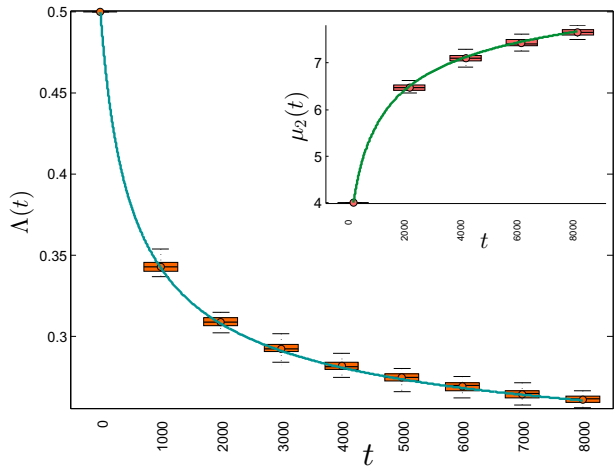}
\caption[Figure ]%
{ (Color online) Temporal evolution of the epidemic threshold for the SIS model. The initial substrate is a ring with 1000 nodes. 
The growth parameters are $\beta=1$ and $\theta=1.1$. 
The error bars pertain to 1000 Monte Carlo simulations. The solid line represents the theoretical prediction~\eqref{th_sis_t}.
The inset shows  the theoretical prediction for the second moment of the degree distribution.
As a result of the addition of new incoming nodes to the network, the epidemic threshold diminishes, making the network more susceptible to endemic outbreaks. 
}
\label{Z_SW_p_sefr_N0_500_T_8000_beta_1_th_1_d_1}
\end{figure}

Here we present more simulation results for the SIS model. 
Another   topology to which~\eqref{th_sis} is applicable is a regular graph. Consider a ring of 500 nodes. Every node would have degree 2, and~\eqref{th_sis} would give the value of 0.5 for the epidemic threshold. Let us study the evolution of the epidemic threshold as new nodes are introduced. We consider the growth parameters $\beta=1$ and $\theta=1.1$. The results are depicted in Figure~\ref{Z_SW_p_sefr_N0_500_T_8000_beta_1_th_1_d_1}. In this case, we observe a uniform increase in the epidemic threshold, which means that the incoming nodes enhance the epidemic resilience of the network right from the inception  of the growth process.

\begin{figure}[h]
\centering
\includegraphics[width=.95 \columnwidth]{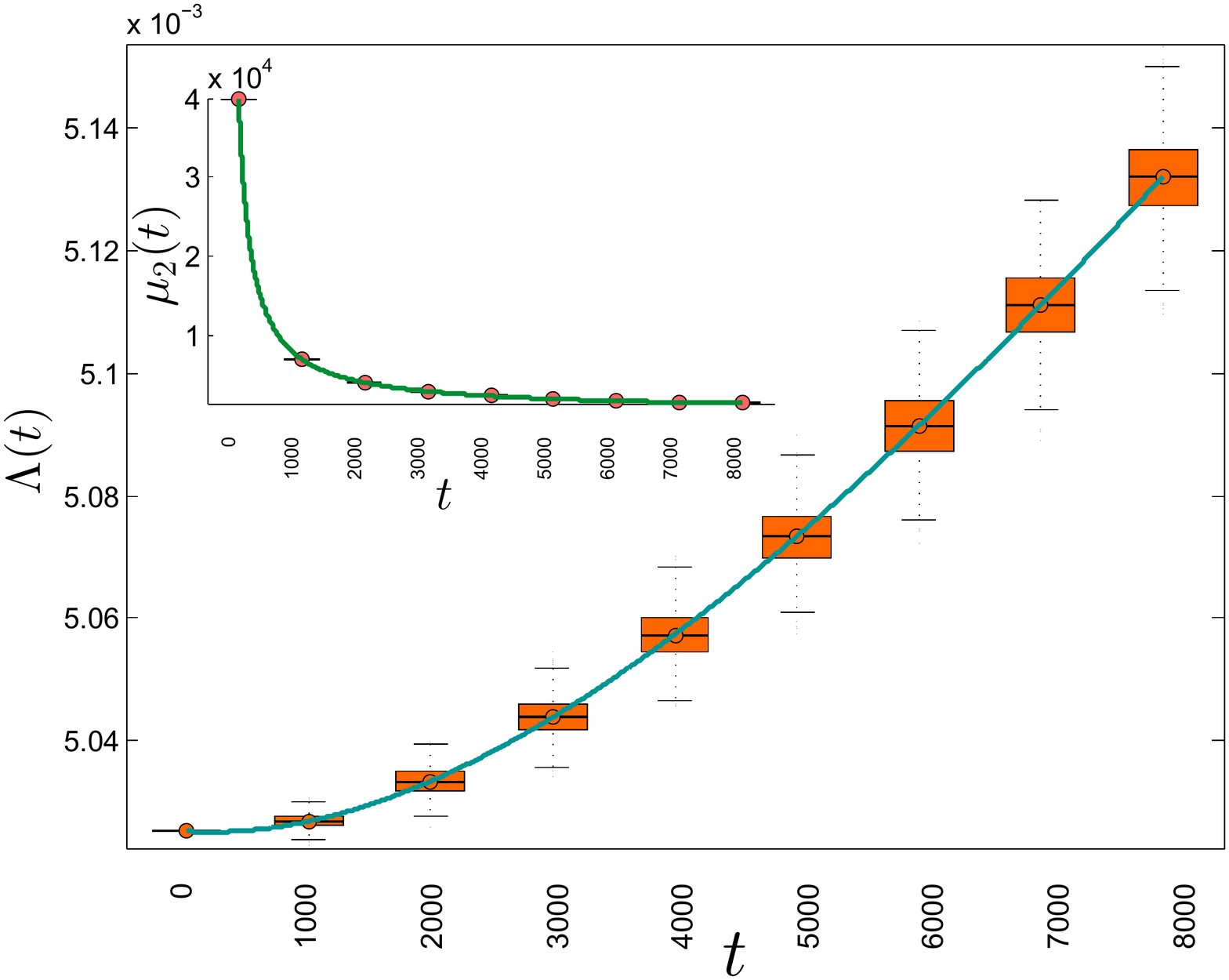}
\caption[Figure ]%
{ (Color online) Temporal evolution of the epidemic threshold for the SIS model. The initial substrate is a  complete graph with 200 nodes. 
The growth parameters are $\beta=1$ and $\theta=1.1$. 
The error bars pertain to 1000 Monte Carlo simulations. The solid line represents the theoretical prediction~\eqref{th_sis_t}.
The inset shows  the theoretical prediction for the second moment of the degree distribution.
As a result of the addition of new incoming nodes to the network, the epidemic threshold diminishes, making the network more susceptible to endemic outbreaks. 
}
\label{ZZZZZSIR_Complete_graph_N0_201_T_8000_beta_1_th_1}
\end{figure}

A special case of a regular graph is the complete graph, in which every node is connected to every other node. Figure~\ref{ZZZZZSIR_Complete_graph_N0_201_T_8000_beta_1_th_1} depicts the temporal evolution of $\Lambda$ for a complete graph with 200 nodes. The growth parameters are $\beta=1$ and $\theta=1.1$. Unlike the previous case (which was a ring), in this case, we observe that the epidemic threshold increases as a result of the arrival of new nodes. Note that the growth parameters of the two settings are identical. This disparity means that in addition to the growth parameters, the topology of the initial network also affects the temporal evolution of the epidemic properties.

\begin{figure}[h]
\centering
\includegraphics[width=.95 \columnwidth]{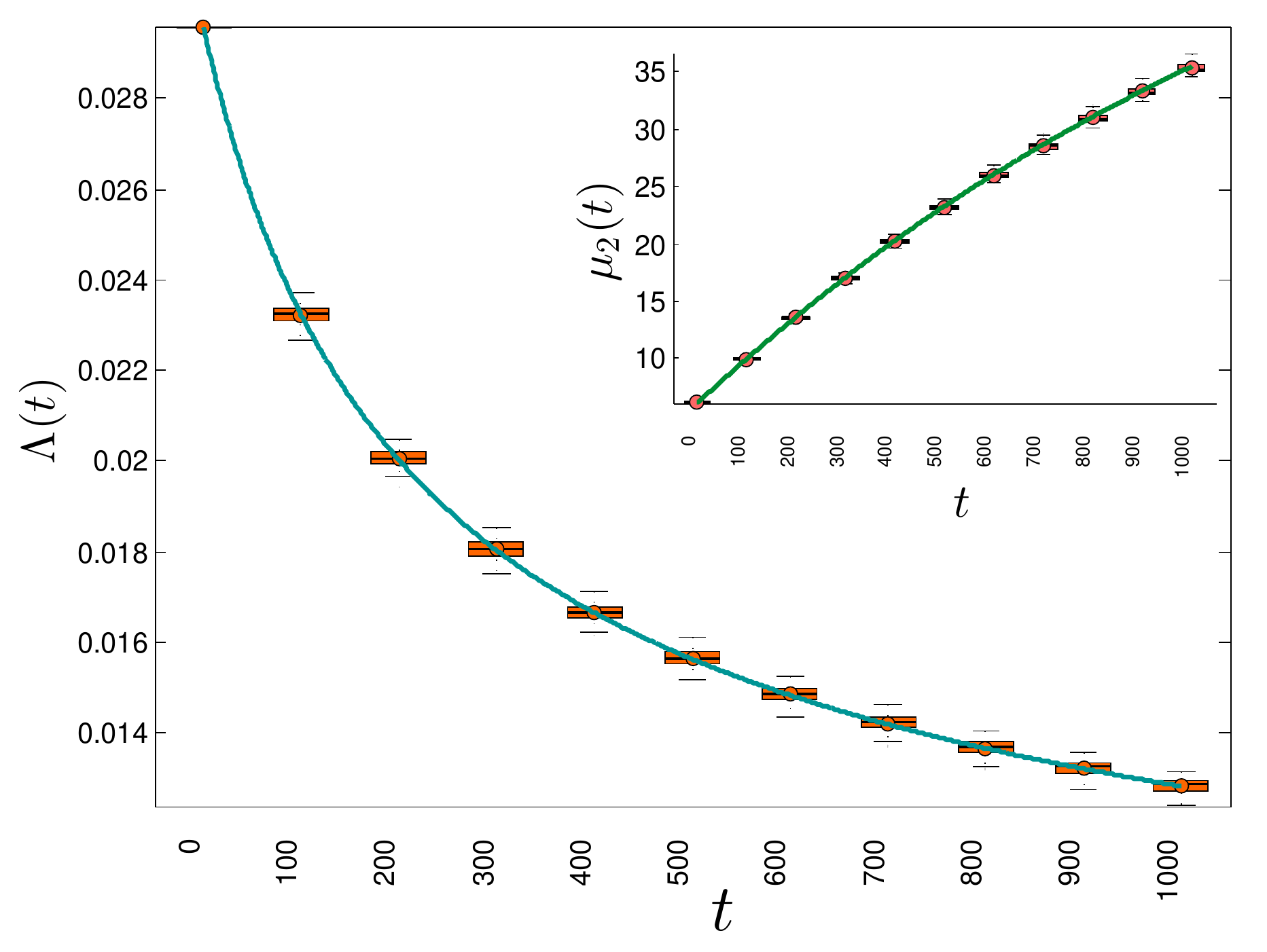}
\caption[Figure ]%
{ (Color online) Temporal evolution of the epidemic threshold for the SIS model. The initial substrate is a   random recursive tree with 1000 nodes. 
The growth parameters are $\beta=4$ and $\theta=0$. 
The error bars pertain to 1000 Monte Carlo simulations. The solid line represents the theoretical prediction~\eqref{th_sis_t}.
The inset shows  the theoretical prediction for the second moment of the degree distribution.
}
\label{Z_SIS_RRT_N0_1000_beta_4_th_0}
\end{figure}

Finally, let us consider the random recursive tree (see, for example, \cite{dorogovtsev2008transition}) which is constructed as follows. Consider a single node. Then introduce new nodes successively to the network, and let each incoming node connect to one existing node chosen uniformly at random.  The randomness renders the network uncorrelated. for simulation purposes, we consider a random recursive tree with 1000 nodes as  the substrate. 
The growth parameters are $\beta=4$ and $\theta=0$. In other words, the growth mechanism coincides with the conventional preferential attachment scheme of the BA model. The results are presented in Figure~\ref{Z_SIS_RRT_N0_1000_beta_4_th_0}. The epidemic threshold decreases uniformly with time, and the simulation results are in good agreement with theoretical predictions.

%

%
%
%
\end{document}